\documentclass[showpacs,preprintnumbers,amsmath,amssymb]{revtex4}
\usepackage{amsmath,amssymb,graphics,epsfig,subfigure}
\usepackage{color}

\begin{document}
\renewcommand{\baselinestretch}{1.3}

\title{Dynamic properties of thermodynamic phase transition for five-dimensional neutral Gauss-Bonnet AdS black hole on free energy landscape}

\author{Shao-Wen Wei \footnote{weishw@lzu.edu.cn},
  Yu-Xiao Liu \footnote{liuyx@lzu.edu.cn},
  Yong-Qiang Wang \footnote{yqwang@lzu.edu.cn, Corresponding author}}

\affiliation{Lanzhou Center for Theoretical Physics, Key Laboratory of Theoretical Physics of Gansu Province, School of Physical Science and Technology, Lanzhou University, Lanzhou 730000, People's Republic of China,\\
 Institute of Theoretical Physics $\&$ Research Center of Gravitation,
Lanzhou University, Lanzhou 730000, People's Republic of China,\\
 Academy of Plateau Science and Sustainability, Qinghai Normal University, Xining 810016, P. R. China}

\begin{abstract}
Understanding the dynamic process of the thermodynamic phase transition can provide the deep insight into the black hole microscopic properties and structures. We in this paper study the dynamic properties of the stable small-large black hole phase transition for the five-dimensional neutral Gauss-Bonnet AdS black hole. Firstly, by using the first law of black holes, we prove that the extremal points of the free energy on the landscape denote the real black hole solutions satisfying the field equations. The local maximal and minimal points correspond to local unstable and stable black hole states, respectively. Especially, on the free energy landscape, the wells of the coexistence small and large black holes have the same depth. Then we investigate the probability evolution governed by the Fokker-Planck equation. Due to the thermal fluctuation, we find that the small (large) black hole state can transit to the large (small) black hole state. Furthermore, the first passage time is calculated. For each temperature, a single peak is presented, which suggests that there is a considerable fraction of the first passage events taking place at short time. And the higher the temperature is, the faster decrease of the probability is. These results will uncover some intriguing dynamic properties of the stable small-large black hole phase transition in modified gravity.
\end{abstract}



\maketitle

\section{Introduction}

Since the discovery of the four laws of the black hole thermodynamics \cite{Hawking0,Bekensteina0,Bekensteinb0,Bardeen0}, black hole phase transitions and microstructures have been one of the significant active areas. Especially, in the anti-de Sitter space, black hole phase transition has attracted much more attention. It was found that there exists a Hawking-Page phase transition between a stable large black hole and a pure thermal radiation \cite{Hawking}. Utilizing the anti-de Sitter/conformal field theory (AdS/CFT) correspondence \cite{Maldacena,Gubser,Witten}, such phase transition was explained as the confinement/deconfinement phase transition of a gauge field \cite{Hawking,Witten2}. Later, the study of the phase transition was generalized to other AdS black holes, and more interesting results were obtained.

In the canonical ensemble, a small-large black hole phase transition was found in charged or rotating AdS black hole backgrounds \cite{Chamblin,Chamblin2,Caldarelli}, which is reminiscent of a liquid-gas transition of the van der Waals (VdW) fluid. Recently, the study was generalized to the extended phase space, where the cosmological constant was interpreted as a pressure \cite{Kastor}. Its corresponding conjugate quantity was treated as the thermodynamic volume of the system and some properties of it were investigated \cite{Dolan00,Cvetic}. In particular, the precise analogy between the small-large black hole phase transition and the liquid-gas phase transition was implemented by Kubiznak and Mann \cite{Kubiznak}. For low temperature, there exists an oscillatory behavior on each isothermal curve, which indicates there is a first-order phase transition between the small and large black holes. Such phase transition extends with the temperature and ends at a critical point. Near that point, critical phenomena were observed \cite{Kubiznak,Samanta17,Samanta16}. Subsequently, much more interesting phase transitions and phase structures were discovered.

Among the studies of the phase transition, probing the black hole microstructures is a hot issue. The black hole systems were found to exhibit a liquid-gas like phase transition of the VdW fluid indicating they share the similar microscopic properties. However, since black hole entropy is proportional to its area rather than the volume, this makes black hole microstructures more mysterious. There are different theories and models to understand black hole entropy. In string theory, the Bekenstein-Hawking entropy area formula was obtained by counting the state number of a weakly coupled D-brane system for supersymmetric black holes \cite{Vafa}. The fuzzball theory suggests that a black hole is constituted by small strings. Then employing the geometry, the entropy formula was derived \cite{Lunin,Mathur}.

As we know, thermodynamic phase transition is a result of the competition between the microcomponents of the system. So the phase transition can reflect some microscopic properties of the black hole, and can help us to peek into the black hole microstructures. Basing on this idea, we developed a possible approach to investigate the black hole microstructures \cite{WeiLiu} following the Ruppeiner geometry \cite{Ruppeiner}. After introducing the number density of the speculative black hole molecules, we first examined its phase structure and showed that there exists a huge change of the black hole microstructures among the first-order phase transition. Then taking the mass and pressure as the fluctuation coordinates, we constructed the Ruppeiner geometry for the charged AdS black holes. Through the corresponding curvature scalar, it indicates that there might be an attractive or repulsive interaction between these molecules. This approach provides us a promising way to test the black hole microstructures, and it was quickly generalized to other AdS black hole backgrounds \cite{Wei:2020cqn,Dehyadegari,Moumni,Sheykhi,Miao,Lis,Chen,Guo,Sheykhi:2019vzb,Xuz,GhoshBhamidipati}.

Basing on the conventional construction of the Ruppeiner geometry and the fluctuation theory, a further step was made in Ref. \cite{Wei2019}. Since for the charged AdS black hole, its heat capacity always vanishes. To cure this problem, we introduced a new normalized scalar curvature. Employing this concept and the empirical interpretation of the geometry, we found that the attractive interaction dominates the most range of the parameters among the black hole molecules. However, there still exists a narrow window for the repulsive interaction for the small black hole of high temperature. This uncovers the unique nature of black hole from the VdW fluid. The critical phenomena of the curvature scalar were also observed. This geometric approach was also generalized to other black hole backgrounds, and many novel properties of the black hole microstructures were disclosed, see Refs. \cite{Wei2019dd,Kumara2020,Yerra:2020oph,Wu:2020fij,
Dehyadegari:2020ebz,Kumara,Mannw,Wei2020d,Wei2020a} for examples.

Accompanying by the study of the black hole phase transitions and phase structures, the dynamic process of the phase transition is a challenge. The mechanism that motivates these black hole molecules transit from one phase to another one is still unknown. Very recently, Li and Wang proposed a way to investigate the dynamic process of the phase transition on the free energy landscape \cite{Li}. In order to reveal the underling kinetics of the phase transition, it was suggested that the dynamics of the phase transition under certain thermal fluctuations can be tested through the associated probabilistic Fokker-Planck equation on the free energy landscape \cite{Zwanzig,Lee,Stell,Wangs,Wolynes}.

The dynamic process for the Hawking-Page phase transition was studied in Einstein gravity and in massive gravity \cite{Li}. Via the Fokker-Planck equation, it was found that the system can change one phase or state to another one between the stable large black hole and the pure thermal radiation under the thermal fluctuations. The first passage process was also calculated. This study showed a novel way to examine the kinetics of the black hole phase transition on the free energy landscape topography. This approach was also generalized to the small-large black hole phase transitions for the charged AdS black hole in Ref. \cite{LiWang}. By numerically calculating the Fokker-Planck equation, they found that the small black hole state can switch to the large black hole state, and the reverse process can also occur. The influence of the temperature and the barrier heights of the free energy landscape were also discussed.

On the free energy landscape, we note that these two wells corresponding to small and large black holes have different depths \cite{LiWang}. That means these two black holes have different Gibbs free energies. According to that a thermodynamic system prefers a low free energy, the case considered in Ref. \cite{LiWang} was actually a phase transition between a stable state and a metastable state. Therefore, the stable small-large black hole phase transition should take place at the case that these two wells share the same depth. This is one of our motivations. On the other hand, as noted before, on the free energy landscape, these transitory states are not real black hole solutions of the field equations. How to judge whether a point on this landscape is real or not is an interesting problem. In this paper, we will start with the first law of black holes, and show a universal result that the extremal points of the corresponding Gibbs free energy denote the real black hole solutions. Meanwhile, the local maximal or minimal point indicates local unstable or stable.

Gauss-Bonnet (GB) gravity is one of the famous modified gravities. In particular, recent study showed a charged GB-AdS black hole in four dimensions by rescaling the GB coupling parameter \cite{Glavan}. Both in canonical and grand canonical ensembles, there exhibits the small and large black hole phase transition \cite{Wei2020d}. The black hole phase transition and instability in GB gravity has been exploited in Refs. \cite{Caijhep2013,Weiprd2014,Zouprd2014,Moprd2015,Konoplyaprd,KonoplyaJHEP}. An attractive property is that there exists the analytical curve of the coexistence small and large black holes for the five-dimensional neutral GB AdS black holes, which will provide us an exact test for the phase transition. In order to disclose the particular preliminary dynamic property of the GB-AdS black hole, we will consider the behaviors of the dynamic process of the small-large black hole phase transition for nonvanishing GB coupling constant in this paper. It is also worthwhile pointing out that the dynamical behavior of the phase transition for the charged GB-AdS black hole has been studied in Ref. \cite{Liwa} after our original study for this neutral GB AdS black holes.

The present paper is organized as follows. In Sec. \ref{tapt}, we briefly review the thermodynamics and phase transition for the five-dimensional neutral Gauss-Bonnet AdS black holes. We show that the change of the horizon radius among the small and large black holes can act as an order parameter to describe the phase transition. Thus the horizon radius provides a good coordinate for the Fokker-Planck equation. In Sec. \ref{gfel}, the free energy landscape is studied. The locations of the real black hole solutions are discussed on the landscape. Global and local stabilities are examined through the first law of black holes. Then the dynamic properties of the stable small-large black hole phase transition are investigated in Sec. \ref{dpotpt}. By solving the Fokker-Planck equation, we obtain the probability evolution of the black hole states. The first passage times are also calculated. Finally, we summarize and discuss our results in Sec. \ref{Conclusion}.

\section{Thermodynamics and phase transition}
\label{tapt}

In this section, we would like to give a brief review of the black hole thermodynamics and phase transition, as well as to examine the order parameter to describe the small-large black hole phase transition corresponding to the black hole horizon radius.

\subsection{Thermodynamics}

\begin{figure}
\center{\subfigure[]{\label{Trh}
\includegraphics[width=7cm]{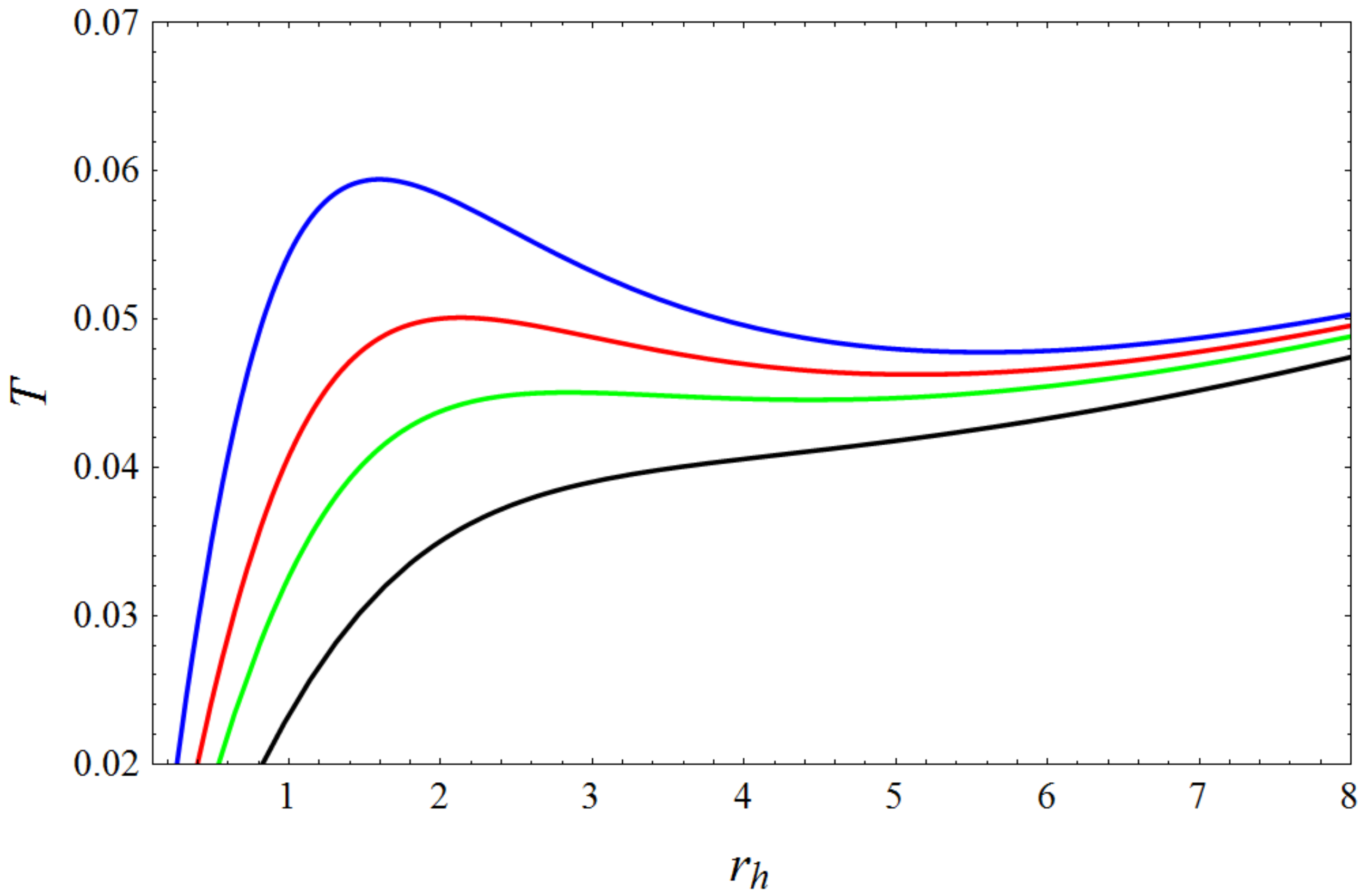}}
\subfigure[]{\label{Arh}
\includegraphics[width=7cm]{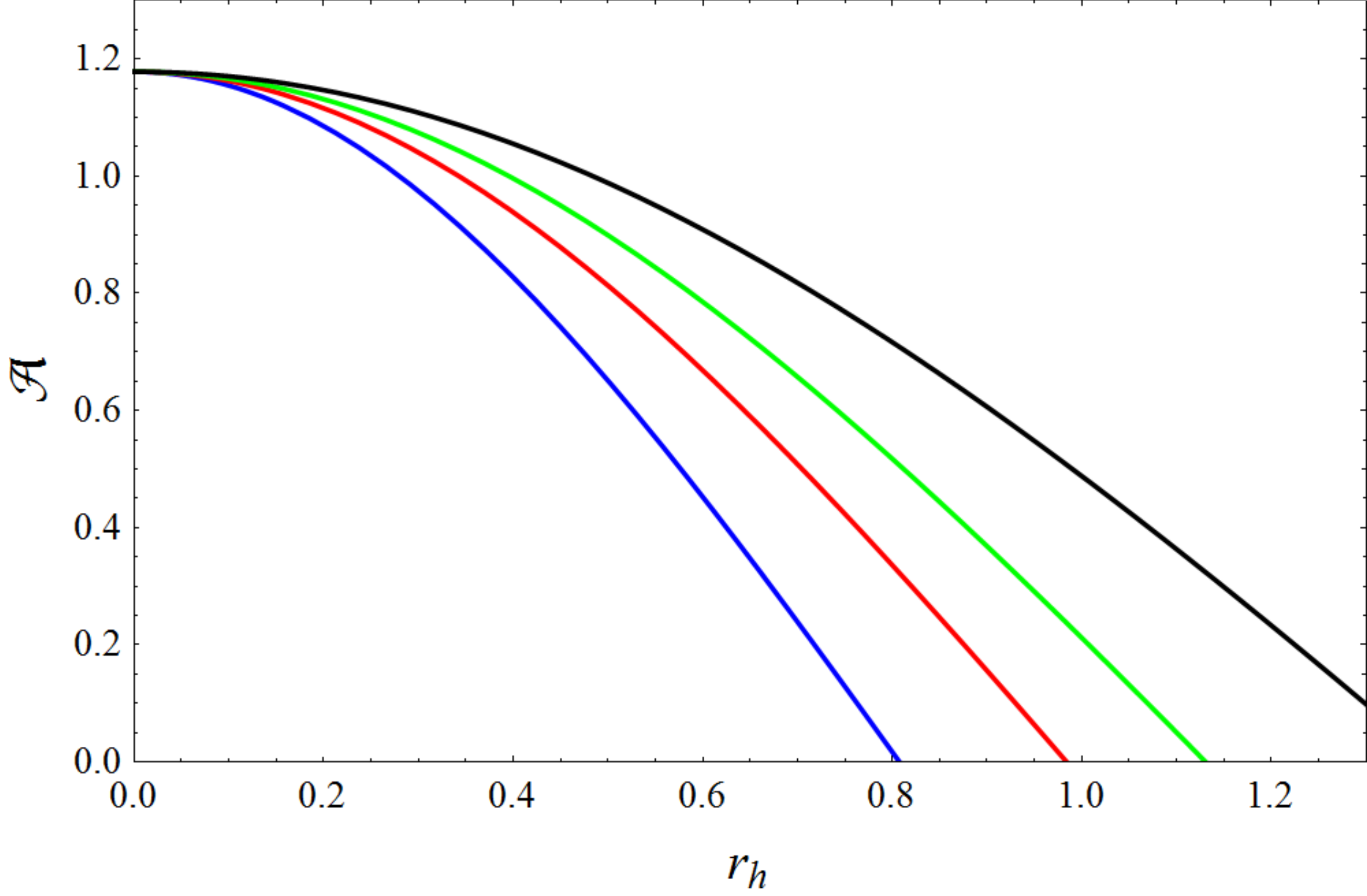}}}
\caption{Behaviors of thermodynamic quantities with $P=0.003$. (a) $T$ vs. $r_{\text{h}}$ with $\alpha$=1, 1.5, 2, and 3 from top to bottom. (b) $\mathcal{A}$ vs. $r_{\text{h}}$ with $\alpha$=1, 1.5, 2, and 3 from bottom to top.}\label{ppArh}
\end{figure}

The action describing a five-dimensional neutral GB-AdS black hole solution is
\begin{eqnarray}
 S=\frac{1}{16\pi G}\int d^{5}x\sqrt{-g}
 \left(\mathcal{R}
    -2\Lambda
    +\alpha_{\text{GB}}\mathcal{L}_{\text{GB}}
  \right), \label{action}
\end{eqnarray}
where $\mathcal{L}_{\text{GB}}=\mathcal{R}_{\mu\nu\gamma\delta}
\mathcal{R}^{\mu\nu\gamma\delta}-4\mathcal{R}_{\mu\nu}\mathcal{R}^{\mu\nu}+\mathcal{R}^{2}$
and $\alpha_{\text{GB}}$ is the GB coupling constant. The solution of the corresponding field equations is
\begin{eqnarray}
 ds^{2}=-f(r)dt^{2}+f^{-1}(r)dr^{2}+r^{2}(d\theta^{2}+\sin^{2}\theta d\phi^{2}+\sin^{2}\theta\sin^{2}\phi d\varphi^2),\label{metric}
\end{eqnarray}
with the metric function given by \cite{Boulware,Cai2,Wiltshire,Cvetic2}
\begin{eqnarray}
 f(r)=1+\frac{r^{2}}{2\alpha}
   \left(1-\sqrt{1 + \frac{32M\alpha}{3\pi r^{4}}-\frac{16P\pi\alpha}{3}}\right).
\end{eqnarray}
where $\alpha=(d-3)(d-4)\alpha_{\text{GB}}$ and the parameter $M$ denotes the black hole mass. In the extended phase space, the cosmological constant $\Lambda$ was interpreted as the thermodynamic pressure \cite{Kastor}
\begin{eqnarray}
 P=-\frac{1}{8\pi}\Lambda.
\end{eqnarray}
The event horizon of the black hole locates at $f(r_{\rm h})=0$. Solving it, one can obtain the black hole mass
\begin{eqnarray}
 M=\frac{\pi}{8}(4P\pi r_{\rm h}^{4}+3r_{\rm h}^{2}+3\alpha).
\end{eqnarray}
According to the thermodynamic law of black holes, the mass should be treated as the enthalpy $H\equiv M$. The temperature, entropy, and thermodynamic volume are
\begin{eqnarray}
 T=\frac{8\pi P r_{\rm h}^{3}+3r_{\rm h}}{6\pi r_{\rm h}^{2}+12\pi\alpha},\quad
 S=\frac{\pi^{2}r_{\rm h}}{2}(r_{\rm h}^{2}+6\alpha), \quad
 V=\frac{\pi^{2}r_{\rm h}^{4}}{2}. \label{EntropyGBBH}
\end{eqnarray}
Moreover, the quantity $\mathcal{A}$ conjugating to the GB coupling coefficient $\alpha$ is obtained
$\mathcal{A}=-\frac{\pi}{8}\frac{32\pi P r_{\rm h}^{4}+9r_{\rm h}^{2}-6\alpha}{r_{\rm h}^{2}+2\alpha}$. We plot $T$ and $\mathcal{A}$ in Fig. \ref{ppArh} with pressure $P$=0.003. For small $\alpha$, the temperature $T$ has a nonmonotonic behavior as $r_\text{h}$. While for large $\alpha$, $T$ increases with $r_\text{h}$. From Fig. \ref{Arh}, we observe that, with the increase of $\alpha$, $\mathcal{A}$ is shifted towards to large $r_\text{h}$.

Moreover, it is easy to check that the following first law and Smarr relation hold
\begin{eqnarray}
 dH&=&TdS+VdP+\mathcal{A}d\alpha,\label{fistlaw}\\
 2H&=&3TS-2PV+2\mathcal{A}\alpha.
\end{eqnarray}
Solving the temperature given in (\ref{EntropyGBBH}), the equation of state for the black hole system is
\begin{eqnarray}
 P=\frac{3 \left(2 \pi  T
   r_{\rm h}^2-r_{\rm h}+4 \pi  \alpha
   T\right)}{8 \pi  r_{\rm h}^3}.
\end{eqnarray}
The equation of state admits a small-large black hole phase transition, which is reminiscent of the liquid-gas phase transition of the VdW fluid. The critical point determined by $(\partial_{r_{\rm h}}P)_{T}=(\partial_{r_{\rm h},r_{\rm h}}P)_{T}=0$ is
\begin{eqnarray}
 P_{\rm c}=\frac{1}{48\pi\alpha},\quad
 T_{\rm c}=\frac{1}{2\pi\sqrt{6\alpha}},\quad
 V_{\rm c}=18\pi^{2}\alpha^{2},\quad
 r_{\rm hc}=\sqrt{6\alpha}.
\end{eqnarray}
Obviously, the critical thermodynamic quantities depend on the GB coupling $\alpha$.

\subsection{Equal area law}

As we know, the critical point is a second-order phase transition. On the other hand, the first-order phase transition can be obtained by constructing the equal areas along each isothermal curve. We present a brief discussion on the equal area law in the following.

\begin{figure}
\center{\subfigure[]{\label{Hort}
\includegraphics[width=7cm]{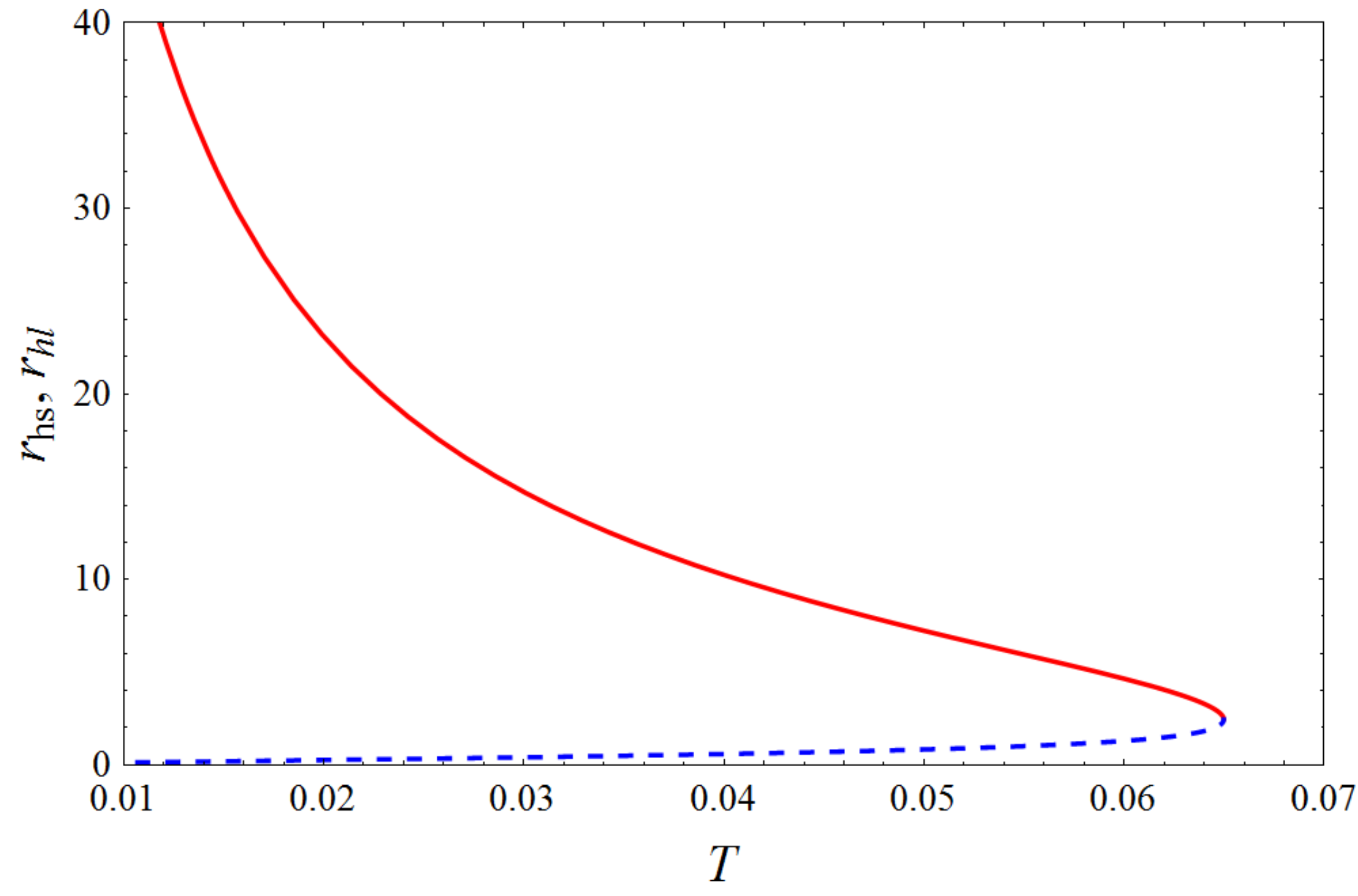}}
\subfigure[]{\label{Drht}
\includegraphics[width=7cm]{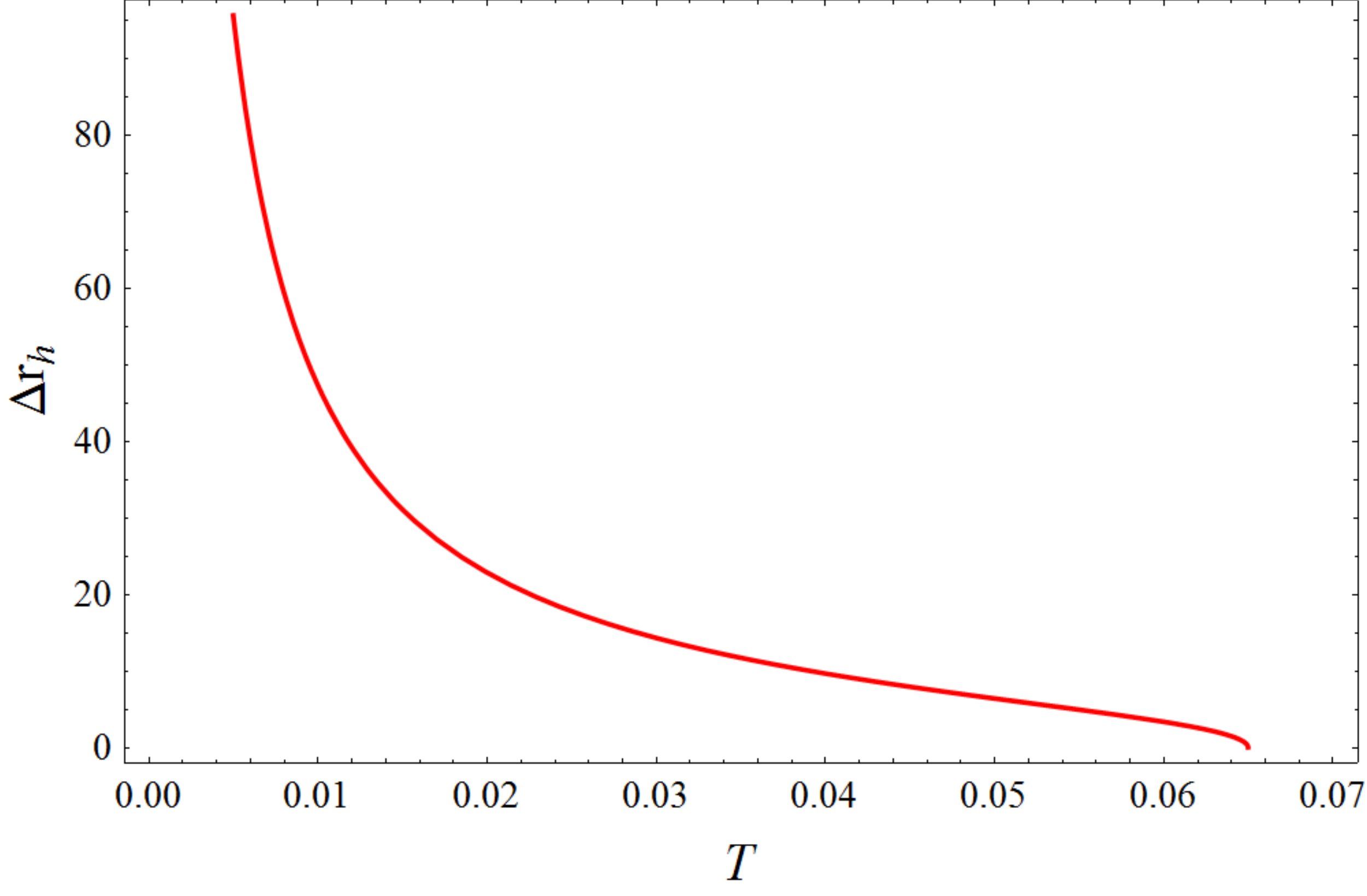}}\\
\subfigure[]{\label{Horizonsl}
\includegraphics[width=7cm]{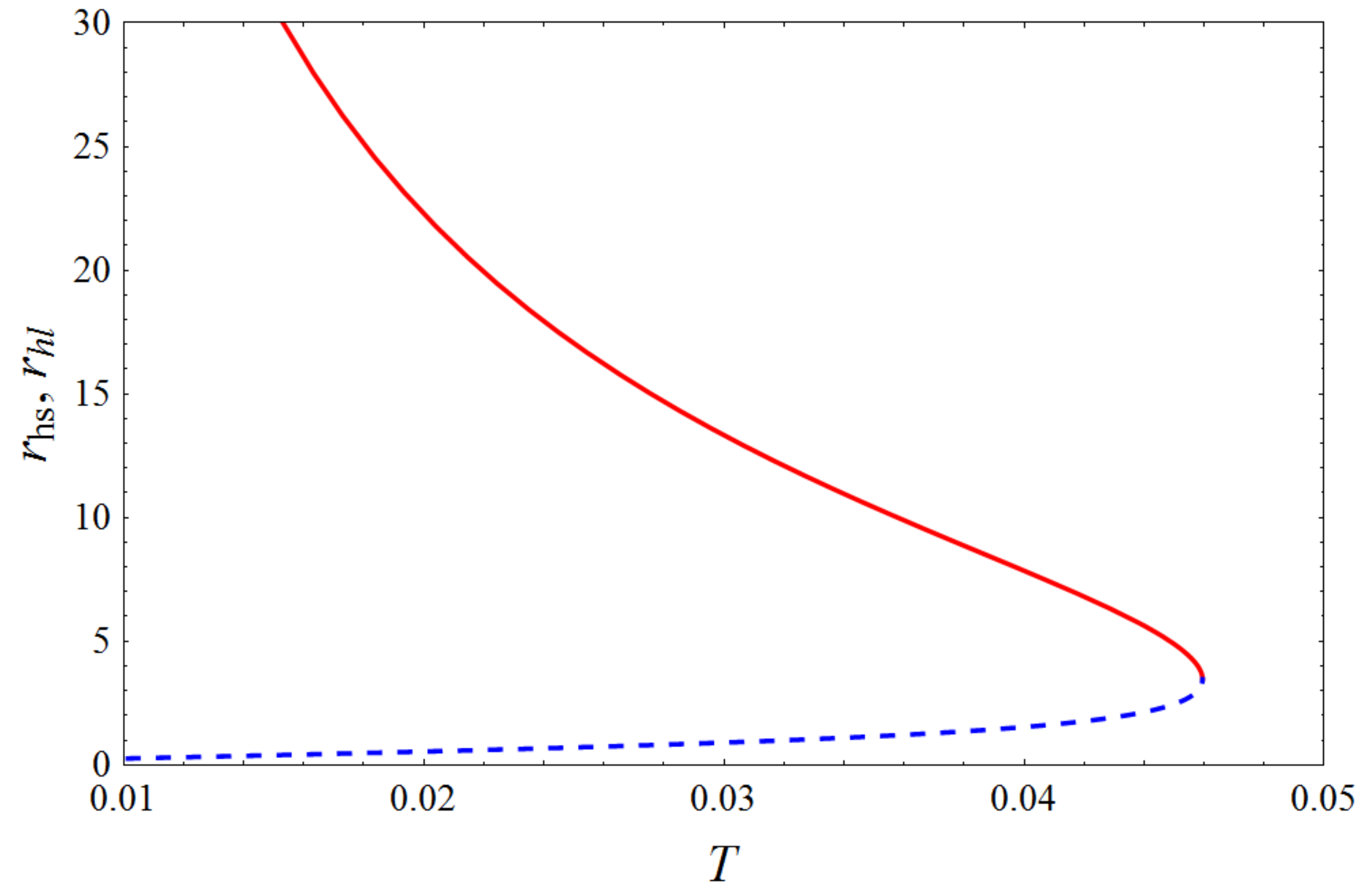}}
\subfigure[]{\label{Charhorizon}
\includegraphics[width=7cm]{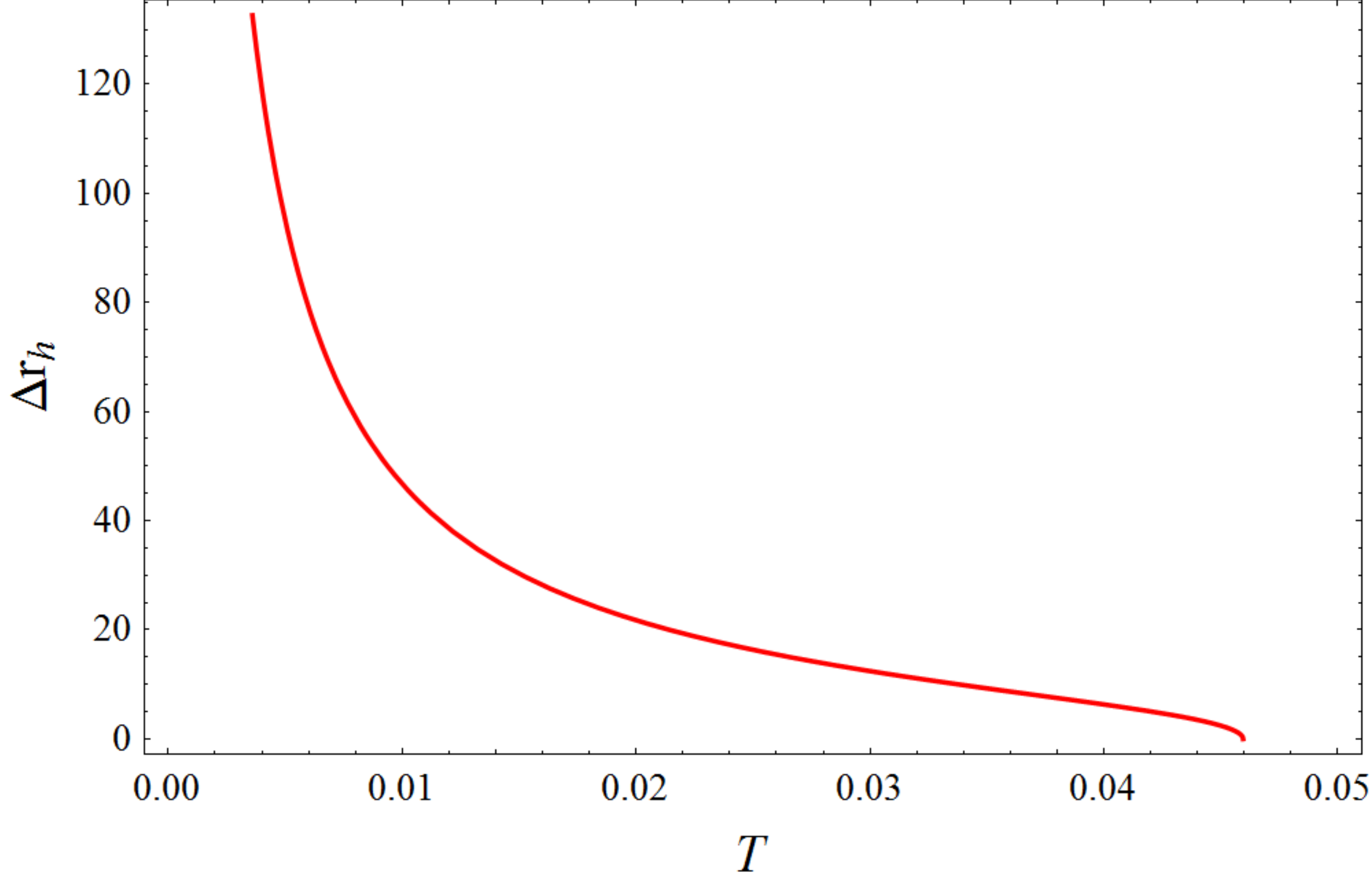}}}
\caption{Behavior of the black hole horizon radius among the phase transition against the temperature. (a) The radii of the coexistence small (blue dashed line) and large (red solid line) black hole horizons with $\alpha$=1. (b) The change of the black hole horizon as a function of the phase transition temperature with $\alpha$=1. (c) The radii of the coexistence small (blue dashed line) and large (red solid line) black hole horizons with $\alpha$=2. (d) The change of the black hole horizon as a function of the phase transition temperature with $\alpha$=2.}\label{ppCharhorizon}
\end{figure}

In terms of the thermodynamic volume, the equation of state can be expressed as
\begin{eqnarray}
 P=\frac{3\pi^{3/2}\alpha T}{2\times(2V)^{3/4}}
 +\frac{3\sqrt{\pi}T}{4\times(2V)^{1/4}}
 -\frac{3}{8\sqrt{2V}},\label{pvsta}
\end{eqnarray}
As we know, the small-large black hole phase transition can be obtained by constructing the Maxwell's equal area law. Labeling the phase transition point with the temperature $T^*$ and pressure $P^*$, the equal area law reads
\begin{eqnarray}
 \int_{V_{\rm s}}^{V_{\rm l}}P(T^*, V)dV=P^*(V_{\rm l}-V_{\rm s}).
\end{eqnarray}
It is worth to point out that the volume here should be the thermodynamic volume rather than the special volume \cite{WeiClapeyron}. Performing the integral, we have
\begin{eqnarray}
 \frac{\left(\sqrt[4]{V_{\rm l}}+\sqrt[4]{V_s}\right) \left(8 P^*
   \left(\sqrt{V_{\rm l}}+\sqrt{V_{\rm s}}\right)+3 \sqrt{2}\right)}{6 \pi \alpha
   +\sqrt{2} \left(\sqrt[4]{V_{\rm l}
   V_{\rm s}}+\sqrt{V_{\rm l}}+\sqrt{V_{\rm s}}\right)}=4
   \sqrt[4]{2} \sqrt{\pi } T^*,\label{pvst0}
\end{eqnarray}
where $V_{\rm l}$ and $V_{\rm s}$ are the thermodynamic volumes of the coexistence large and small black holes. Moreover, for these two coexistence phases, the equation of state holds
\begin{eqnarray}
 P^*&=&\frac{3\pi^{3/2}\alpha T^*}{2\times(2V_{\rm s})^{3/4}}
 +\frac{3\sqrt{\pi}T^*}{4\times(2V_{\rm s})^{1/4}}
 -\frac{3}{8\sqrt{2V_{\rm s}}},\label{pvsta}\\
 P^*&=&\frac{3\pi^{3/2}\alpha T^*}{2\times(2V_{\rm l})^{3/4}}
 +\frac{3\sqrt{\pi}T^*}{4\times(2V_{\rm l})^{1/4}}
 -\frac{3}{8\sqrt{2V_{\rm l}}}.\label{pvstb}
\end{eqnarray}
Combining (\ref{pvst0})-(\ref{pvstb}), one gets the analytical coexistence curve \cite{Moprd2015}
\begin{eqnarray}
 P^*=\frac{3-\sqrt{9-192\pi^2\alpha T^{*2}}}{96\pi \alpha}.
\end{eqnarray}
For a given phase transition temperature $T$, the horizon radius of the coexistence small and large black holes are
\begin{eqnarray}
 r_{\rm hs}=\frac{6\left(4\pi\alpha T-\sqrt{\alpha \left(48\pi^2 \alpha T^2+\sqrt{9-192\pi^2\alpha T^2}-3\right)}\right)}{3-\sqrt{9-192\pi^2\alpha T^2}},\label{rsmall}\\
 r_{\rm hl}=\frac{6\left(4\pi\alpha T+\sqrt{\alpha \left(48\pi^2 \alpha T^2+\sqrt{9-192\pi^2\alpha T^2}-3\right)}\right)}{3-\sqrt{9-192\pi^2\alpha T^2}}.\label{rlarge}
\end{eqnarray}
We describe the behaviors of $r_{\rm hs}$ and $r_{\rm hl}$ in Figs. \ref{Hort} and \ref{Horizonsl} with $\alpha$=1 and 2, respectively. With the increase of the temperature, it is easy to observe that $r_{\rm hs}$ increases while $r_{\rm hl}$ decreases. At the critical temperature, they meet each other. This also implies that there exists a sudden change of the black hole horizon among the phase transition below the critical temperature. We denote the change as
\begin{eqnarray}
 \Delta r_{\rm h}=r_{\rm hl}-r_{\rm hs}
 =\frac{12 \sqrt{\alpha  \left(48 \pi ^2 \alpha
   T^2+\sqrt{9-192 \pi ^2 \alpha
   T^2}-3\right)}}{3-\sqrt{9-192 \pi ^2 \alpha
   T^2}}.
\end{eqnarray}
From Figs. \ref{Horizonsl} and \ref{Charhorizon}, it is clear that $\Delta r_{\rm h}$ is a decreasing function of the temperature for both $\alpha$=1 and 2. For low temperature, $\Delta r_{\rm h}$ takes a finite value. While at the critical temperature, $\Delta r_{\rm h}$ vanishes. Near the critical point, we expand it in the following form
\begin{eqnarray}
 \Delta r_{\rm h}=12\times(24\pi^2\alpha^3)^{\frac{1}{4}}\times(T_{\rm c}-T)^{\frac{1}{2}}
 +\mathcal{O}(T_{\rm c}-T)^{\frac{3}{2}}.
\end{eqnarray}
Obviously, $\Delta r_{\rm h}$ has a critical exponent $\frac{1}{2}$. Therefore, combining these properties of $\Delta r_{\rm h}$, it can serve as an order parameter to characterize the small-large black hole phase transition.

On the other hand, we can also expend the horizon radius (\ref{rsmall}) and (\ref{rlarge}) near the critical temperature after a shift of the critical horizon radius
\begin{eqnarray}
 r_{\text{hc}}-r_{\text{hs}}&=&6\times(24\pi^2\alpha^3)^{\frac{1}{4}}\times(T_{\rm c}-T)^{\frac{1}{2}}
 +\mathcal{O}(T_{\rm c}-T),\\
 r_{\text{hl}}-r_{\text{hc}}&=&6\times(24\pi^2\alpha^3)^{\frac{1}{4}}\times(T_{\rm c}-T)^{\frac{1}{2}}
 +\mathcal{O}(T_{\rm c}-T)
\end{eqnarray}
Obviously, after the shift, the horizon radius of the coexistence small and large black hole share a critical $\frac{1}{2}$. When $T<T_\text{c}$, they are nonzero, while vanishes at the critical temperature. Therefore, the horizon radius can be treated as the order parameter of the phase transition.

\section{Gibbs free energy landscape}
\label{gfel}

Gibbs free energy is an important thermodynamic quantity to investigate the phase transition in addition to the equal area law. For the first-order phase transition, it exhibits a swallow tail behavior. At the second-order phase transition point, the Gibbs free energy is continuous but not smooth. Moreover, it also corresponds to the thermal dynamic phase transition of a black hole.

\begin{figure}
\center{\subfigure[]{\label{Hort}
\includegraphics[width=7cm]{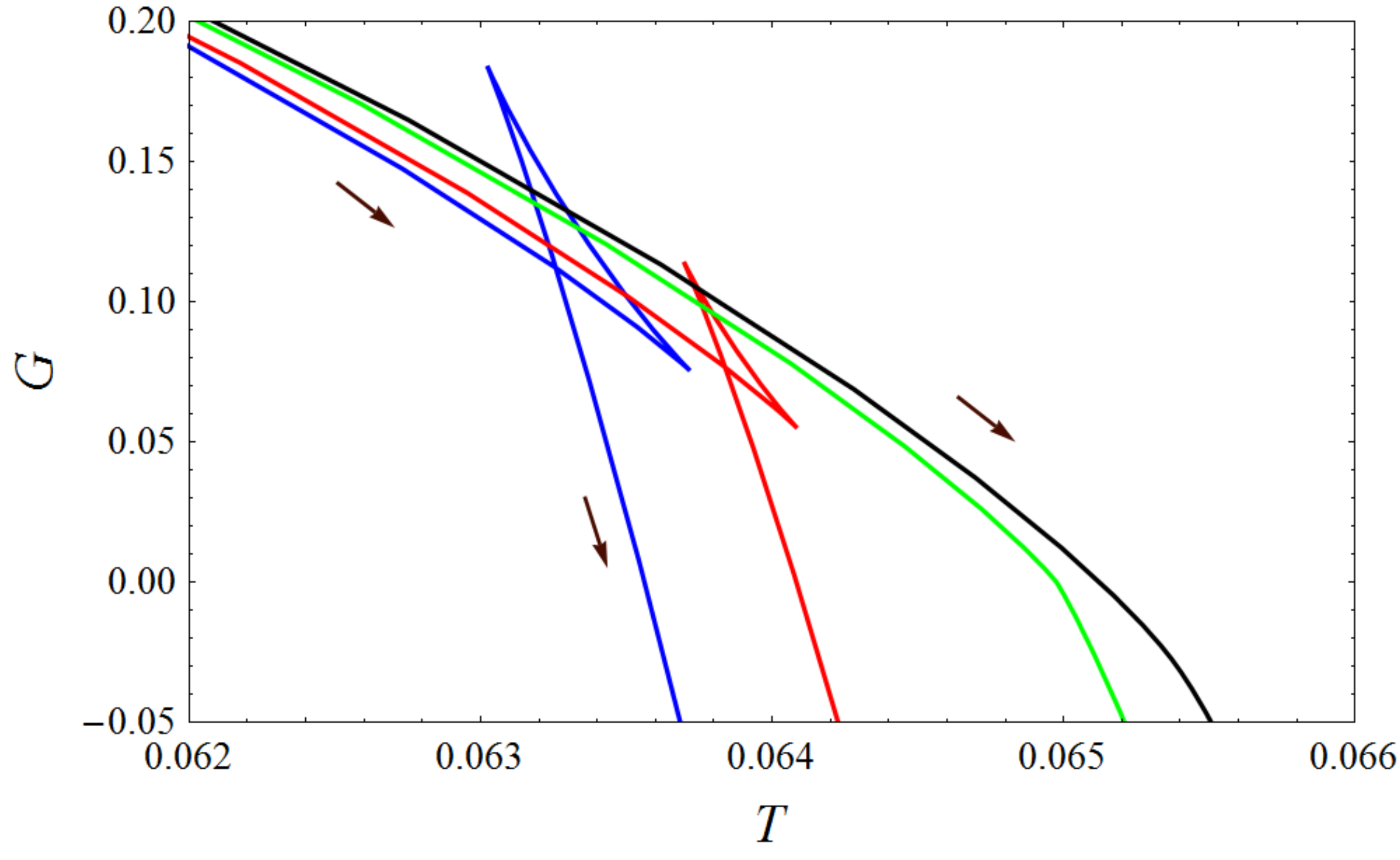}}\subfigure[]{\label{Gibbs}
\includegraphics[width=7cm]{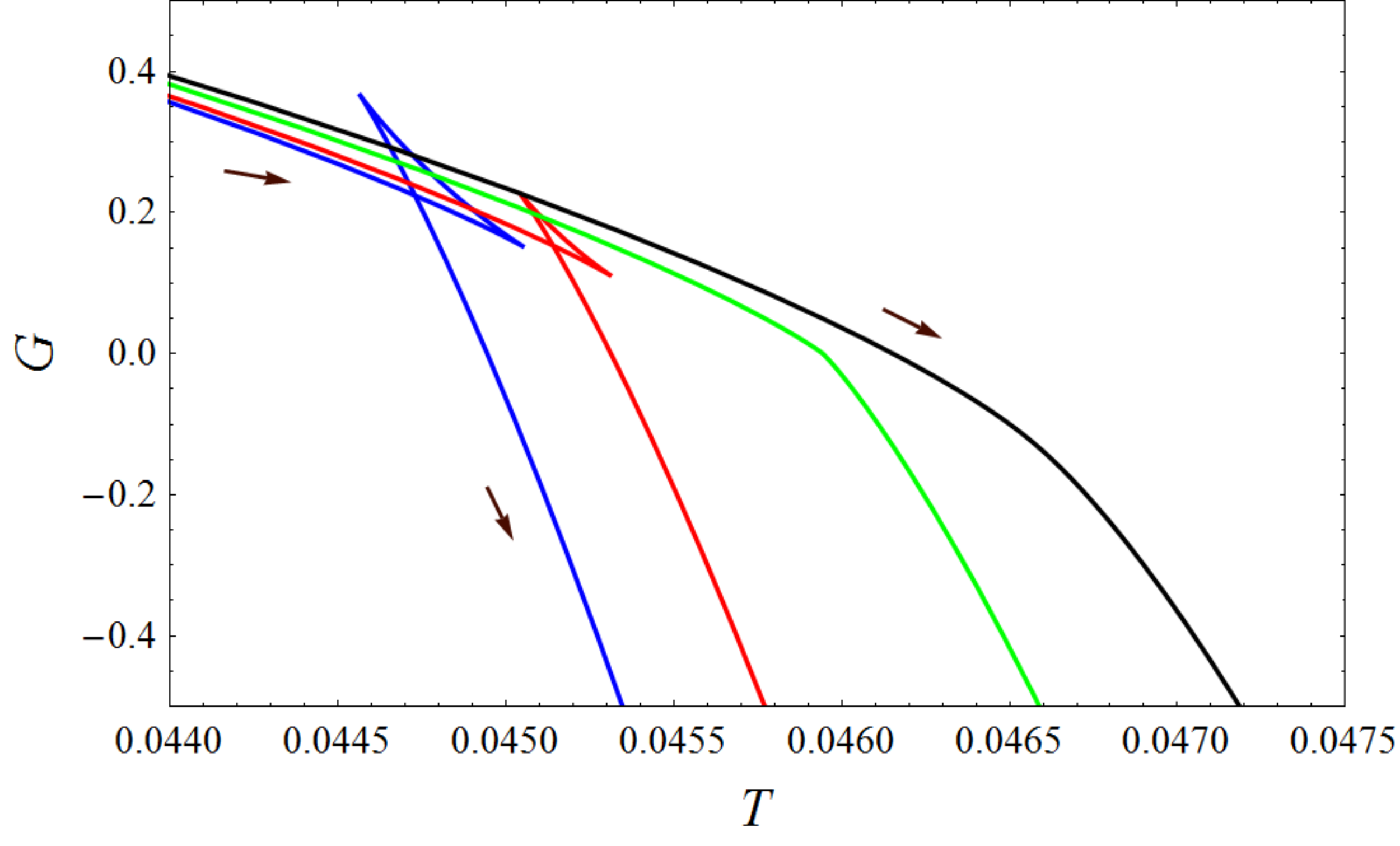}}}
\caption{Gibbs free energy as a function of the temperature. (a) The GB coupling $\alpha$=1 and $P$=0.0060, 0.0062, 0.006631 ($P_{\rm c}$), and 0.0068 from left to right. (b) The GB coupling $\alpha$=2 and the pressure $P$=0.0030, 0.0031, 0.003316 ($P_{\rm c}$), and 0.0035 from left to right. The arrows indicate the increase of $r_{\rm h}$.}\label{ppGibbs}
\end{figure}

For a five-dimensional neutral GB AdS black hole, the Gibbs free energy is
\begin{eqnarray}
 G=H-TS=\frac{\pi\left(18\alpha^2-4\pi Pr_{\rm h}^6+r_{\rm h}^4
   (3-72\pi\alpha P)-9\alpha r_{\rm h}^2\right)}{24\left(2\alpha
   +r_{\rm h}^2\right)}.
\end{eqnarray}
The behavior of $G$ is shown in Fig. \ref{ppGibbs} with $\alpha$=1 and 2, respectively. Although different $\alpha$ gives different critical temperature, we can still find that when the pressure is lower than its critical value, there is a swallow tail behavior. Such behavior disappears beyond the critical point. According to the thermodynamics, a system always prefers a state of lower Gibbs free energy. Therefore, the phase transition occurs at the intersection point of the small and large black hole branches.

On the Gibbs free energy landscape, the free energy $G_{\rm L}$ reads
\begin{eqnarray}
 G_{\rm L}=H-T_{\rm E}S=\frac{1}{8}\pi \left(3\alpha+4\pi P r_{\rm h}^4-4\pi r_{\rm h}
   \left(6 \alpha+r_{\rm h}^2\right) T_{\rm E}+3 r_{\rm h}^2\right).\label{GL}
\end{eqnarray}
Note that $T_{\rm E}$ denotes the temperature of the ensemble rather than the black hole Hawking temperature. As suggested in Refs. \cite{Li,LiWang}, this Gibbs free energy $G_{\rm L}$ only describes a real black hole when $T_{\rm E}=T$. For fixed $P$ and $T_{\rm E}$, one can investigate the behavior of $G_{\rm L}$ a function of $r_{\rm h}$ which corresponds to the order parameter for the Hawking-Page phase transition and small-large black hole phase transition. Here we would like to address this issue clearly with taking $P$=0.006 as an example.

\begin{figure}
\center{\subfigure[]{\label{GibssSkip}
\includegraphics[width=7cm]{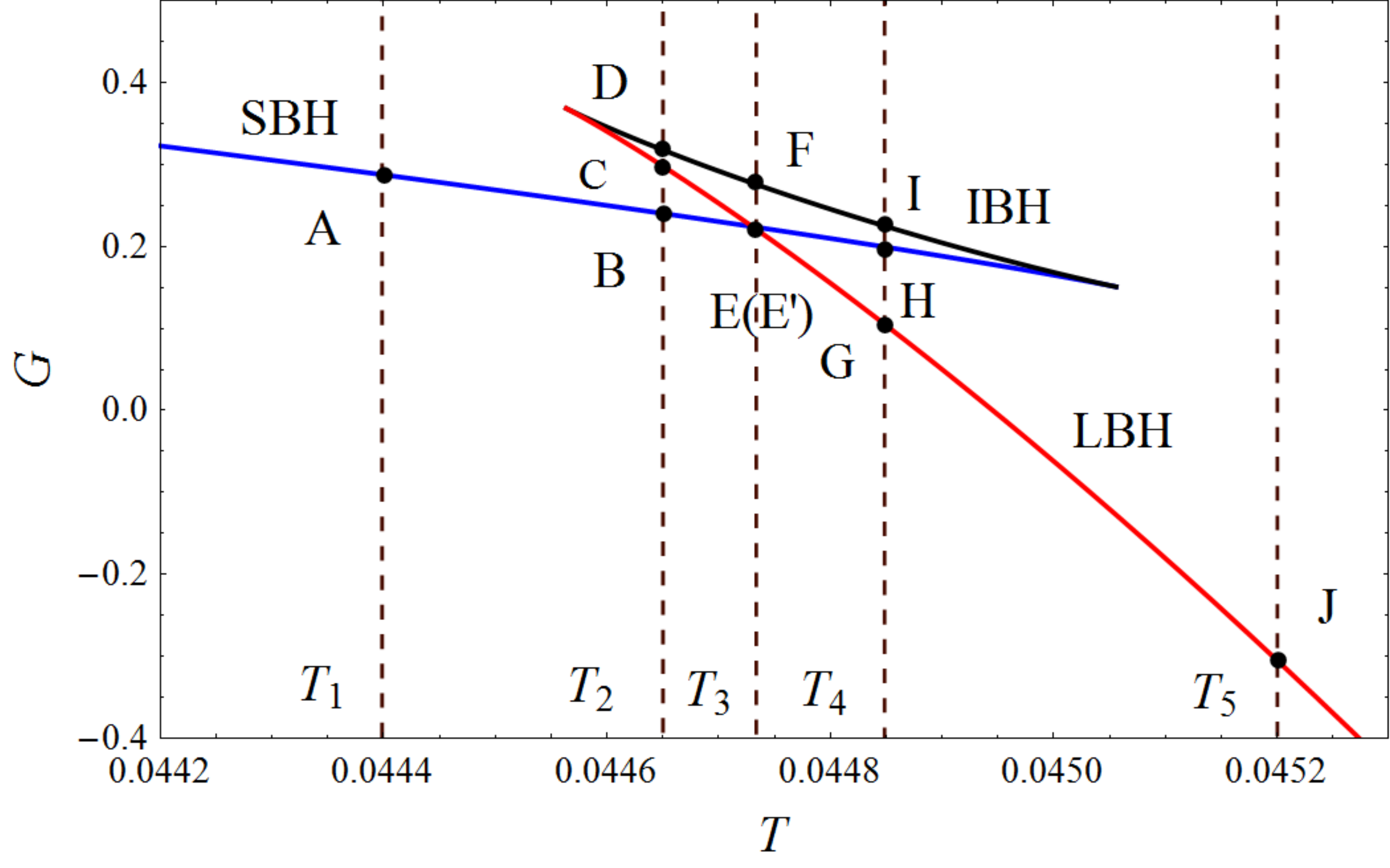}}
\subfigure[]{\label{Gibssgb}
\includegraphics[width=7cm]{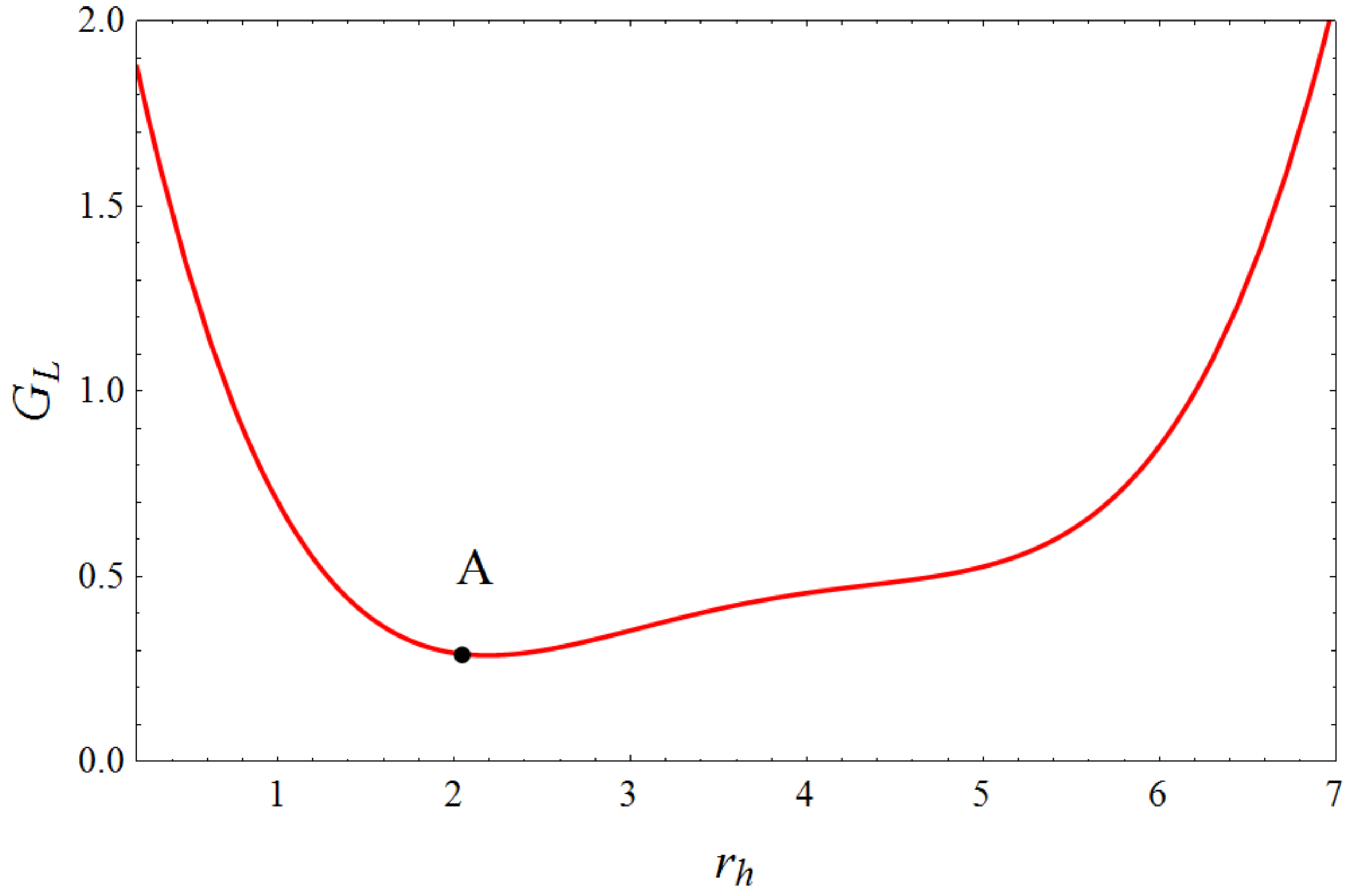}}
\subfigure[]{\label{Gibssgc}
\includegraphics[width=7cm]{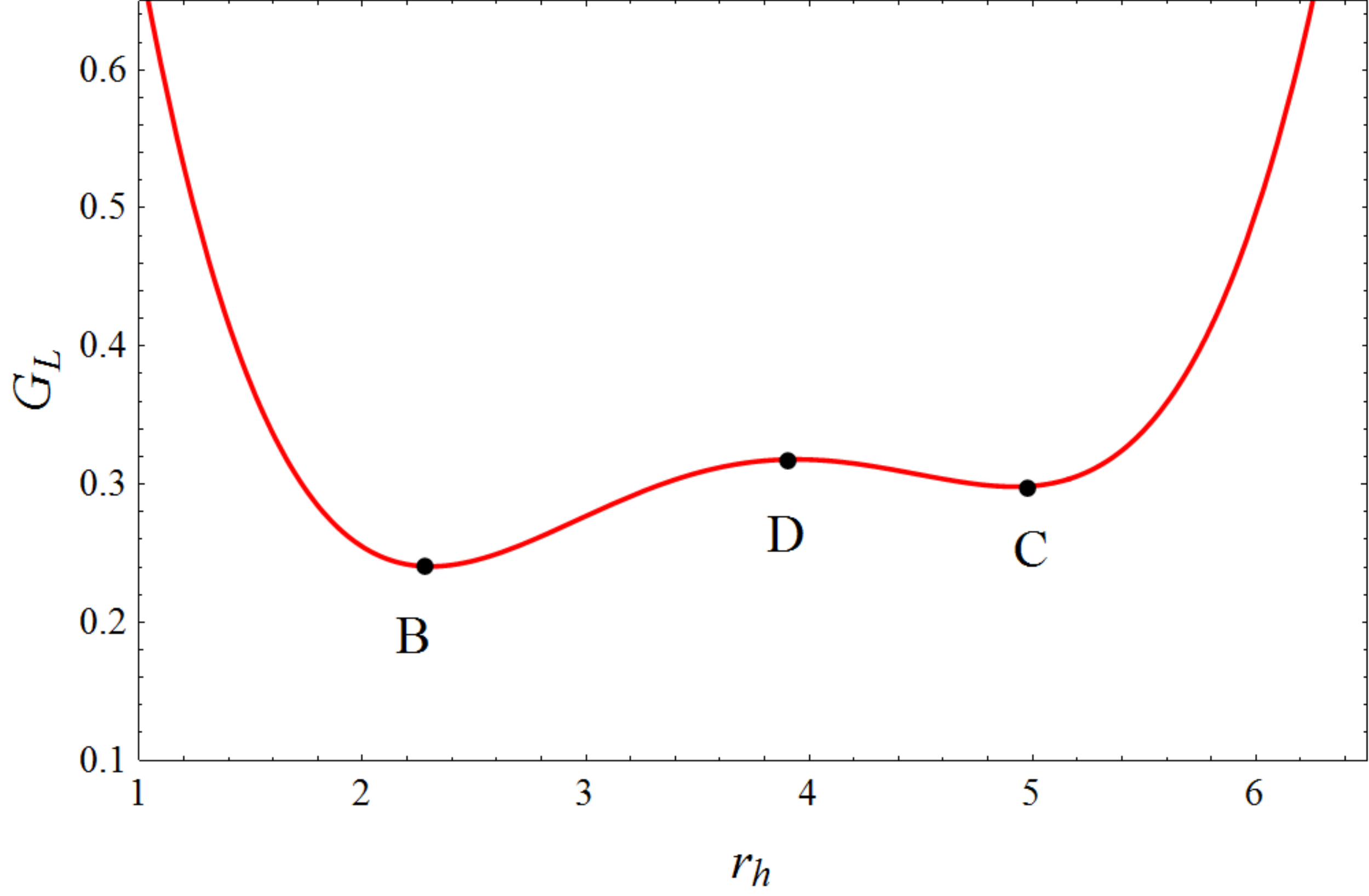}}
\subfigure[]{\label{Gibssgd}
\includegraphics[width=7cm]{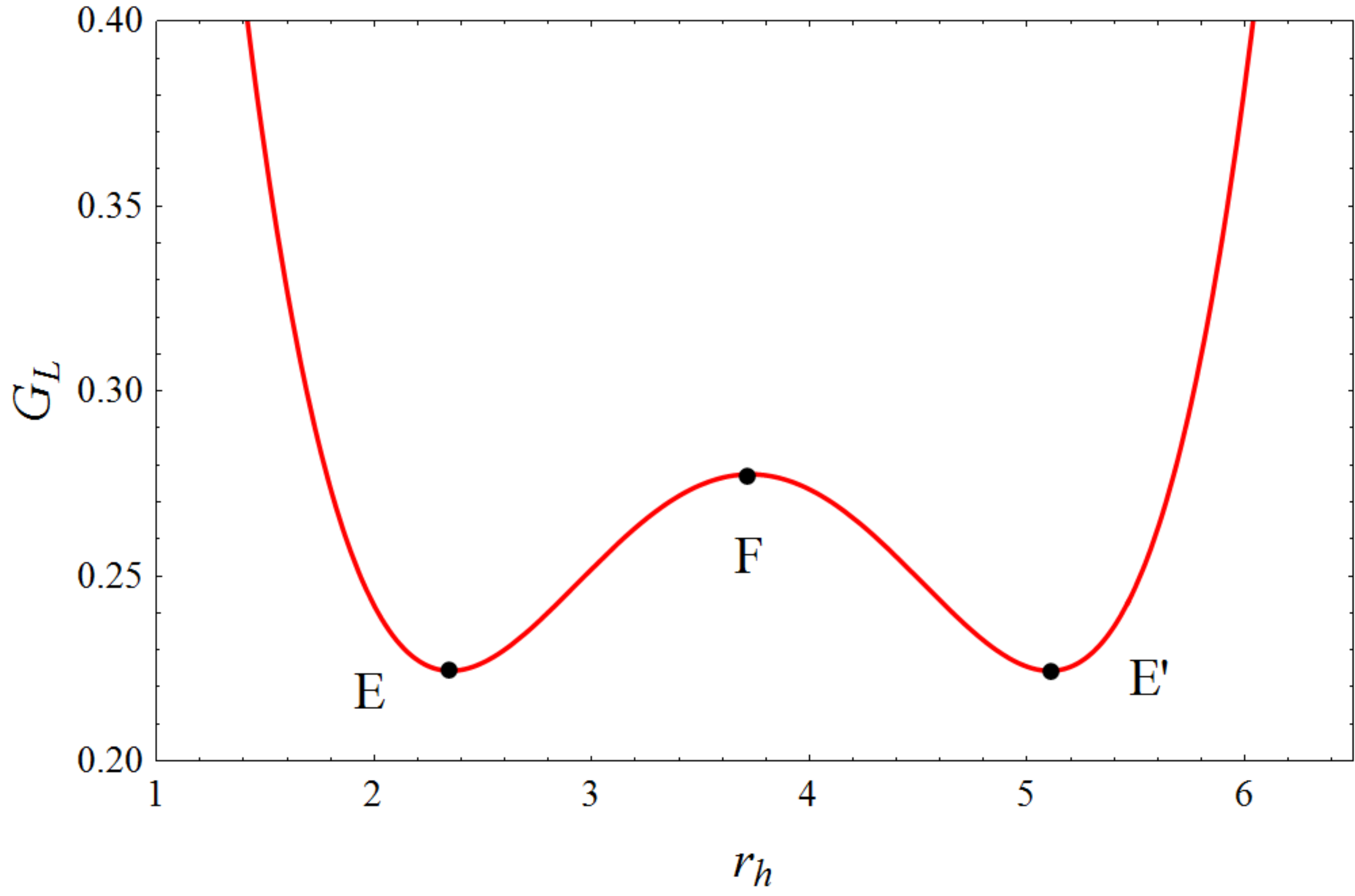}}
\subfigure[]{\label{Gibssge}
\includegraphics[width=7cm]{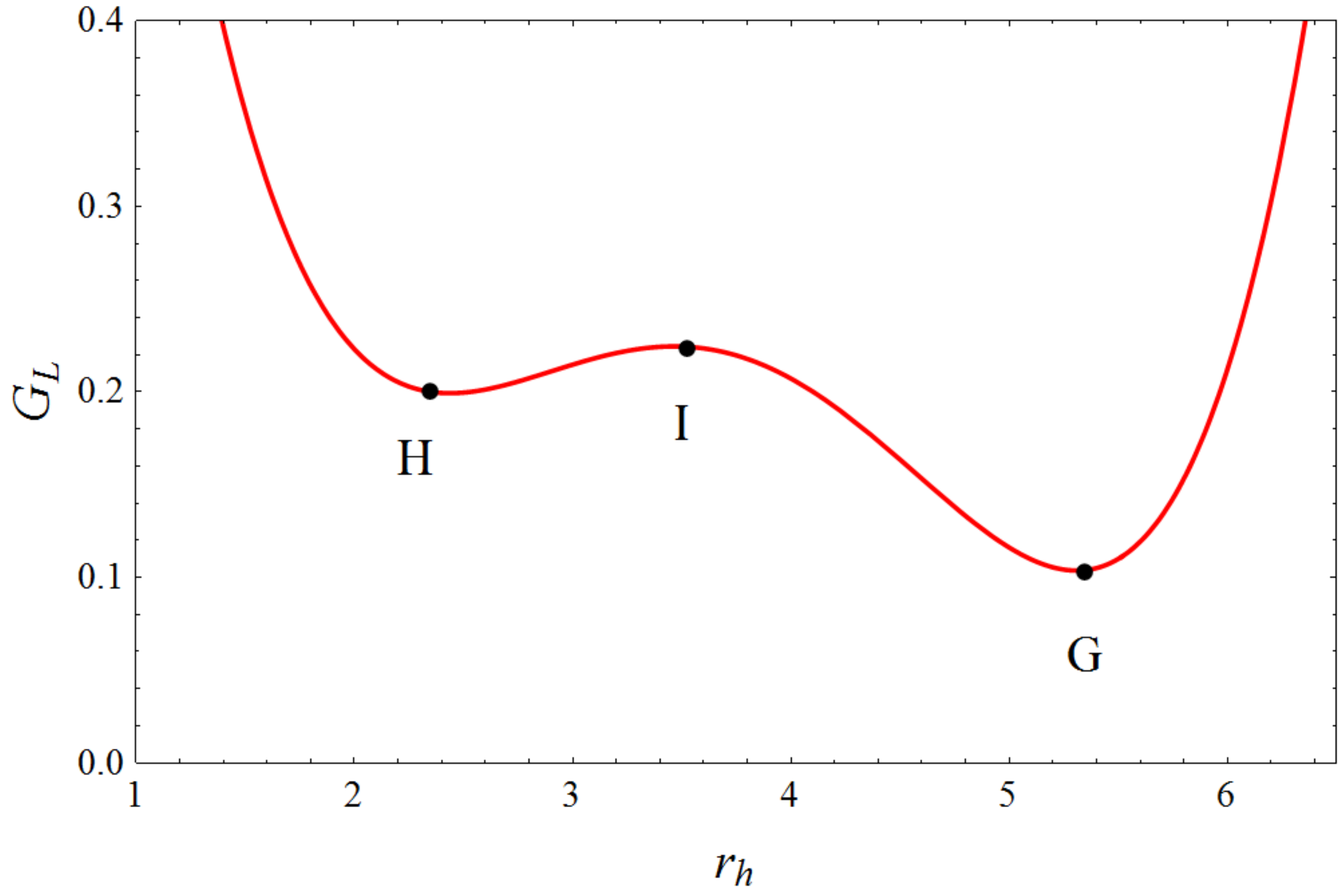}}
\subfigure[]{\label{Gibssgf}
\includegraphics[width=7cm]{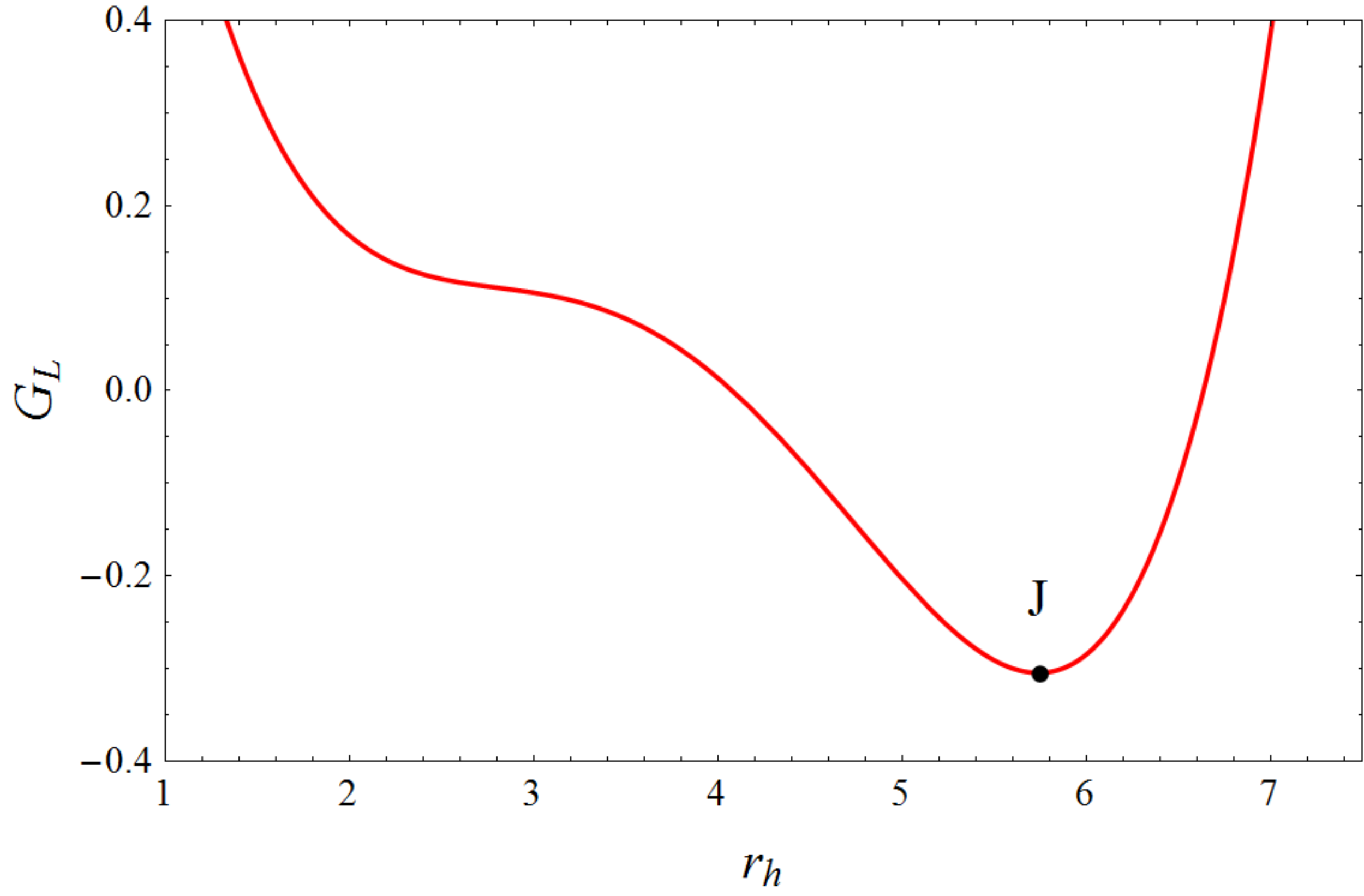}}}
\caption{Behaviors of the Gibbs free energy with $\alpha$=2 and $P$=0.003. (a) $G$ vs $T$. Blue, black, and red lines are for the small black hole (SBH), intermediate black hole (IBH), and large black hole (LBH) branches, respectively. The temperatures $T_1$-$T_5$=0.04440, 0.04465, 0.04473, 0.04485, and 0.04520. (b) $G_{\rm L}$ vs $r_{\rm h}$ with $T_{1}$=0.04440. (c) $G_{\rm L}$ vs $r_{\rm h}$ with $T_{2}$=0.04465. (d) $G_{\rm L}$ vs $r_{\rm h}$ with $T_{3}$=0.04473. (e) $G_{\rm L}$ vs $r_{\rm h}$ with $T_{4}$=0.04485. (f) $G_{\rm L}$ vs $r_{\rm h}$ with $T_{2}$=0.04520.}\label{ppGibssgf}
\end{figure}

First, we display the behavior of $G$ as the temperature $T$ in Fig. \ref{GibssSkip}. The small, large, and intermediate black hole branches are described by the blue, black, and red lines. In particular, the small and large black hole branches have positive heat capacity, and thus they are local stable, while the intermediate black hole branch of negative heat capacity is local unstable. Considering that the system prefers the state of the lowest Gibbs free energy, some states on the small and large black hole branches, see the states H and C, will be global unstable even they have positive heat capacity. With the increase of the temperature, the black hole system goes through states A-B-E-G-J.

Taking the temperature $T$=$T_1$-$T_5$, we plot the Gibbs free energy $G_{\rm L}$ as the horizon radius in Fig. \ref{ppGibssgf}, respectively. For $T$=$T_1$ shown in Fig. \ref{Gibssgb}, the Gibbs free energy has only one extremal point located at state A, which also has the lowest value. From Fig. \ref{GibssSkip}, it is clear that state A belongs to the small black hole branch. When $T$=$T_2$, different from the case $T$=$T_1$, three extremal points are presented. One is the local maximum, which corresponds to the intermediate black hole. Other two are the local minima. Since state B has the lowest Gibbs free energy, the system will prefer the small black hole. And the large black hole described by state C is metastable. Increasing the temperature such that $T$=$T_3$, three extremal points are still given. However, these two local extremal points have the same Gibbs free energy. So, states E and E$'$ are both the real states that the system prefers. States E and E$'$ are the small and large black hole phases, respectively. Hence, this case just describes the small-large black hole phase transition. Or it is a coexistence phase of small and large black holes. Contrary to $T$=$T_2$, the case of $T$=$T_4$ allows the existence of the large black hole described by state G. Meanwhile, state H is a metastable, which is a metastable small black hole. For $T$=$T_5$, the metastable small black hole and intermediate black hole disappear and only the large black hole is allowed. In summary, states A and B are the local and global stable small black holes. States G and J are the local and global large black holes. States D, F, and I are the local and global unstable intermediate black holes. Sates C and H are the local stable while global unstable large and small black holes, respectively. More importantly, states E and E$'$ are the coexistence small and large black holes, and the small-large black hole phase transition occurs between them. In what follows, we only discuss this case, where these two wells have the same depth. Note that in Ref. \cite{LiWang}, the authors considered the case where these two wells have different depths, so actually it is a phase transition between a metastable black hole and a stable black hole instead of a stable small-large black hole phase transition.

In summary, we would like to give some comments for this Gibbs free energy landscape. i) Only the extremal point of $G_{\rm L}$ denotes the real black hole phase, which satisfies the Einstein field equations. The local maximum and minimum correspond to the unstable and stable black hole phases, respectively. ii) For some cases, $G_{\rm L}$ displays the double well behavior. The lowest well corresponds to the global stable black hole phase. While another is global unstable. iii) The small-large black hole phase transition occurs at the case that these wells have the same depth, see Fig. \ref{Gibssgd}. Furthermore, we expect that more rich phase transitions, such as the reentrant phase transition and triple point, take place when more wells appear.

In the following, we attempt to examine the extremal point and stability by using the first law of black holes. By using (\ref{GL}), we have
\begin{eqnarray}
 \left(\frac{\partial G_{\rm L}}{\partial r_{\rm h}}\right)_{P, T_{\rm E}, \alpha}
  &=&\left(\frac{\partial M}{\partial r_{\rm h}}\right)_{P, \alpha}-T_{\rm E}\left(\frac{\partial S}{\partial r_{\rm h}}\right)_{P, \alpha}\nonumber\\
  &=&\left(\frac{\partial M}{\partial S}\right)_{P, \alpha}\left(\frac{\partial S}{\partial r_{\rm h}}\right)_{P, \alpha}-T_{\rm E}\left(\frac{\partial S}{\partial r_{\rm h}}\right)_{P, \alpha}\nonumber\\
  &=&(T-T_{\rm E})\left(\frac{\partial S}{\partial r_{\rm h}}\right)_{P, \alpha},
\end{eqnarray}
where the first law (\ref{fistlaw}) was used in the last step. On the Gibbs free energy landscape, a real black hole locates at the place with $T=T_{\rm E}$, which is obvious at the extremal point of $G_{\rm L}$ satisfying $\left(\frac{\partial G_{\rm L}}{\partial r_{\rm h}}\right)_{P, T_{\rm E}, \alpha}$=0. Next, considering the second derivative of $G_{\rm L}$ to $r_{\rm h}$ at the extremal point, we obtain
\begin{eqnarray}
 \left(\frac{\partial^2 G_{\rm L}}{\partial r_{\rm h}^{2}}\right)_{P, T_{\rm E}, \alpha}\bigg|_{T=T_{\rm E}}
  &=&\left(\frac{\partial T}{\partial r_{\rm h}}\right)_{P, \alpha}\left(\frac{\partial S}{\partial r_{\rm h}}\right)_{P, \alpha}
  +(T-T_{\rm E})\left(\frac{\partial^2 S}{\partial r_{\rm h}^2}\right)_{P, \alpha}\bigg|_{T=T_{\rm E}}\nonumber\\
  &=&\left(\frac{\partial T}{\partial S}\right)_{P, \alpha}\left(\frac{\partial S}{\partial r_{\rm h}}\right)_{P, \alpha}^2\nonumber\\
  &=&\frac{T}{C_{P}}\left(\frac{\partial S}{\partial r_{\rm h}}\right)_{P, \alpha}^2,\label{gg}
\end{eqnarray}
where the heat capacity $C_P=T\left(\frac{\partial S}{\partial T}\right)_{P, \alpha}$. The second term in the first line vanishes due to the extremal condition. For a nonextremal black hole, its temperature is always positive. So (\ref{gg}) indicates that $\left(\frac{\partial^2 G_{\rm L}}{\partial r_{\rm h}^{2}}\right)_{P, T_{\rm E}, \alpha}>0$ (or $<0$) produces $C_P>0$ (or $<0$). Thus on the Gibbs free energy landscape, local maximum or minimum of $G_{\rm L}$ is local thermodynamic unstable or stable, respectively.

\section{Dynamic properties of thermodynamic phase transition}
\label{dpotpt}

In the above section, we have examined the Gibbs free energy landscape for the black hole. The small, intermediate, and large black holes, as well as the phase transitions between them, have been clearly exhibited. In this section, we aim to investigate the dynamic process of the stable small-large black hole phase transition. Since the patterns are similar, we here only focus on the GB coupling parameter $\alpha$=2.

\subsection{Fokker-Planck equation and probabilistic evolution}

As shown above, via the Gibbs free energy landscape, the small-large black hole phase transition takes place at the case, where the double wells have the same depth. In this section, we will examine the probability evolution for the black hole system in this picture.

For certain pressure and temperature, the Gibbs free energy $G_{\rm L}$ (\ref{GL}) is a function of the black hole horizon $r_{\rm h}$. In the following, we denotes $r_{\rm h}$ as $r$ for simplicity. Our purpose is to study the probability evolution of the black hole system under a thermal fluctuation.

The probabilistic evolution of the black hole states is determined by the Fokker-Planck equation via the Gibbs free energy landscape, which reads \cite{Zwanzig,Lee,Stell,Wangs,Wolynes}
\begin{eqnarray}
 \frac{\partial \rho(r, t)}{\partial t}=D
 \frac{\partial}{\partial r}\left(e^{-\beta G_{\rm L}(T, P, r)}
 \frac{\partial}{\partial r}\left(e^{\beta G_{\rm L}(T, P, r)}\rho(r, t)\right)\right),\label{FPeq}
\end{eqnarray}
where the parameter $\beta=\frac{1}{k_{\rm B}T}$ and the diffusion coefficient $D=\frac{k_{\rm B}T}{\zeta}$ with $k_{\rm B}$ and $\zeta$ being the Boltzman constant and dissipation coefficient, respectively. The function $\rho(r, t)$ describes the probability distribution that the black hole prefers after a thermal fluctuation. Without loss of generality, we set $k_{\rm B}$=$\zeta$=1 in the following.

To solve the Fokker-Planck equation, one should impose the boundary conditions and the initial condition. Let us first consider the conditions at the boundary $r=r_0$. There are two types, one is the reflection boundary condition, which preserves the normalization of the probability distribution, and another one is the absorbing boundary condition. They are, respectively, given by
\begin{eqnarray}
 \beta G_{\rm L}'(T, P, r_0)\rho(r_0, t)+\rho'(r_0, t)=0,\\
 \rho(r_0, t)=0,
\end{eqnarray}
where the prime denotes the derivative with respect to $r$. In our following calculation, we first adopt the reflection boundary condition, and the boundaries locate at $r$=0 and $\infty$. The initial condition is chosen as a Gaussian wave packet located at $r_{\rm i}$,
\begin{eqnarray}
 \rho(r, 0)=\frac{1}{\sqrt{\pi}a}e^{-\frac{(r-r_{\rm i})^2}{a^2}}.
\end{eqnarray}
Since we focus on the dynamic process of the phase transition between the small and large black holes, $r_{\rm i}$ can be set to $r_{\rm hs}$ or $r_{\rm hl}$ given in (\ref{rsmall}) or (\ref{rlarge}), which means the black hole system is initially at small or large black hole state. We expect, after some time, there are nonzero probability distributions both at the small and large black hole states, which will indicate that there indeed exists a phase transition between the small and large black holes.

\begin{figure}
\center{\subfigure[]{\label{ThreeDrho44s}
\includegraphics[width=7cm]{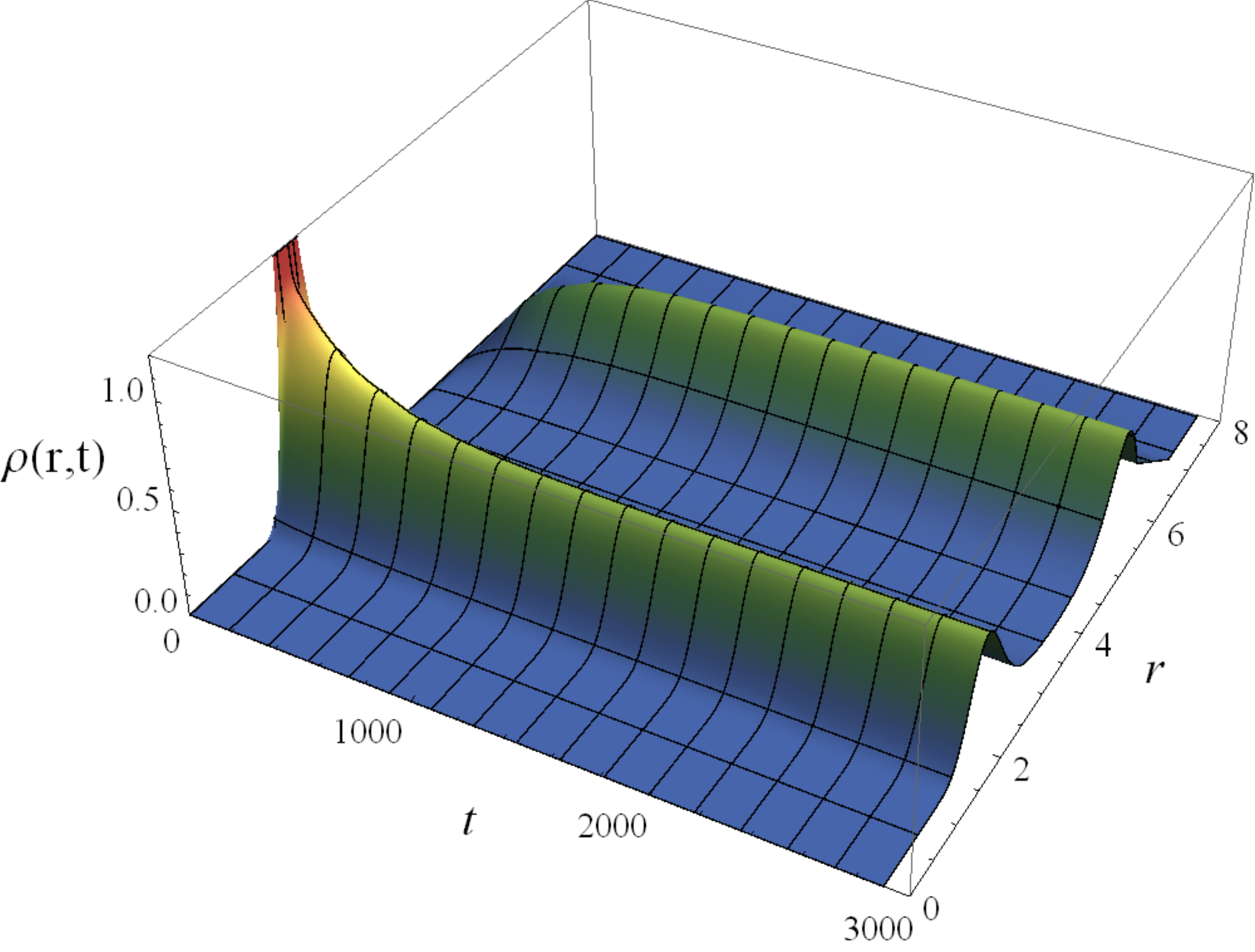}}
\subfigure[]{\label{ThreeDrho45s}
\includegraphics[width=7cm]{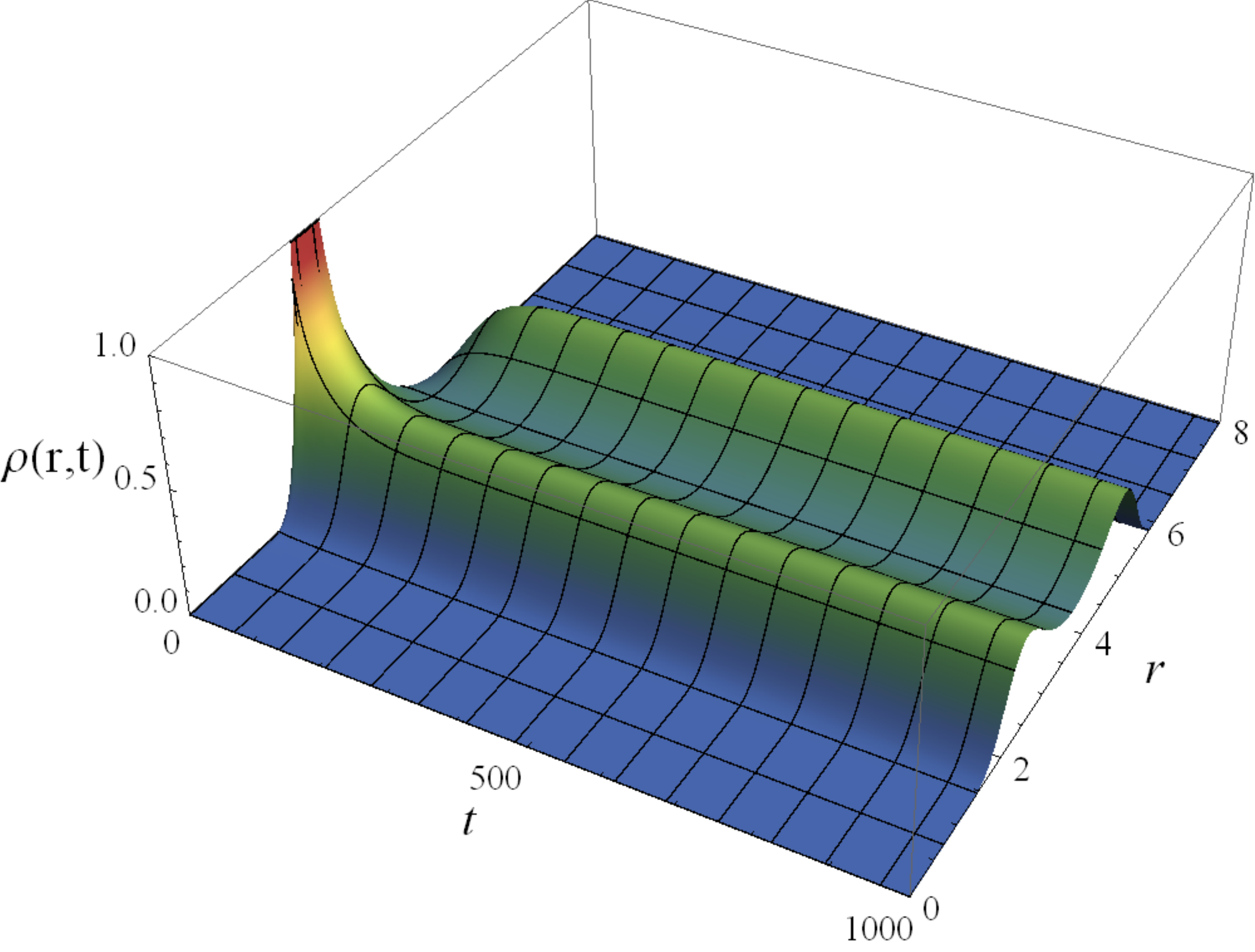}}
\subfigure[]{\label{ThreeDrho44l}
\includegraphics[width=7cm]{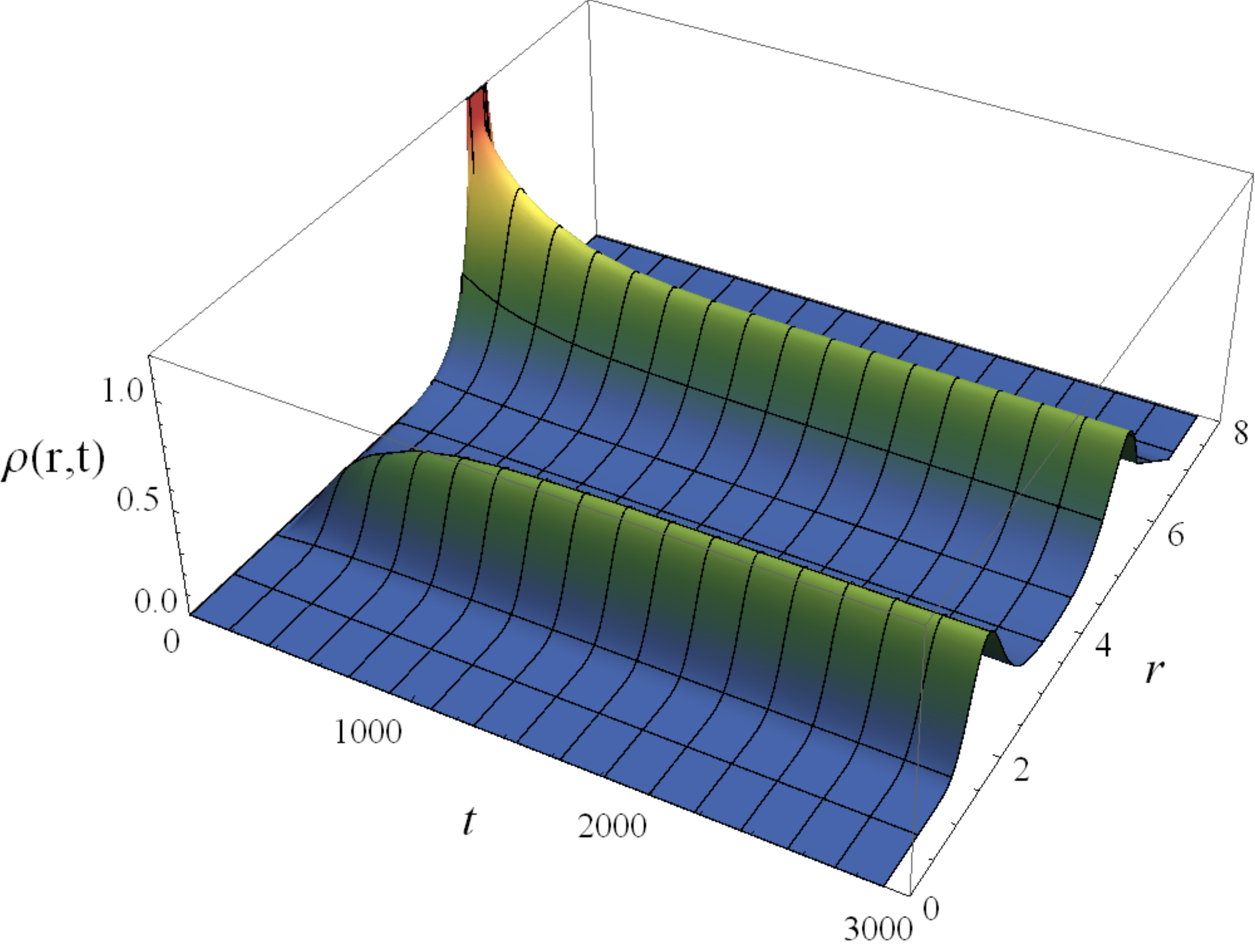}}
\subfigure[]{\label{ThreeDrho45l}
\includegraphics[width=7cm]{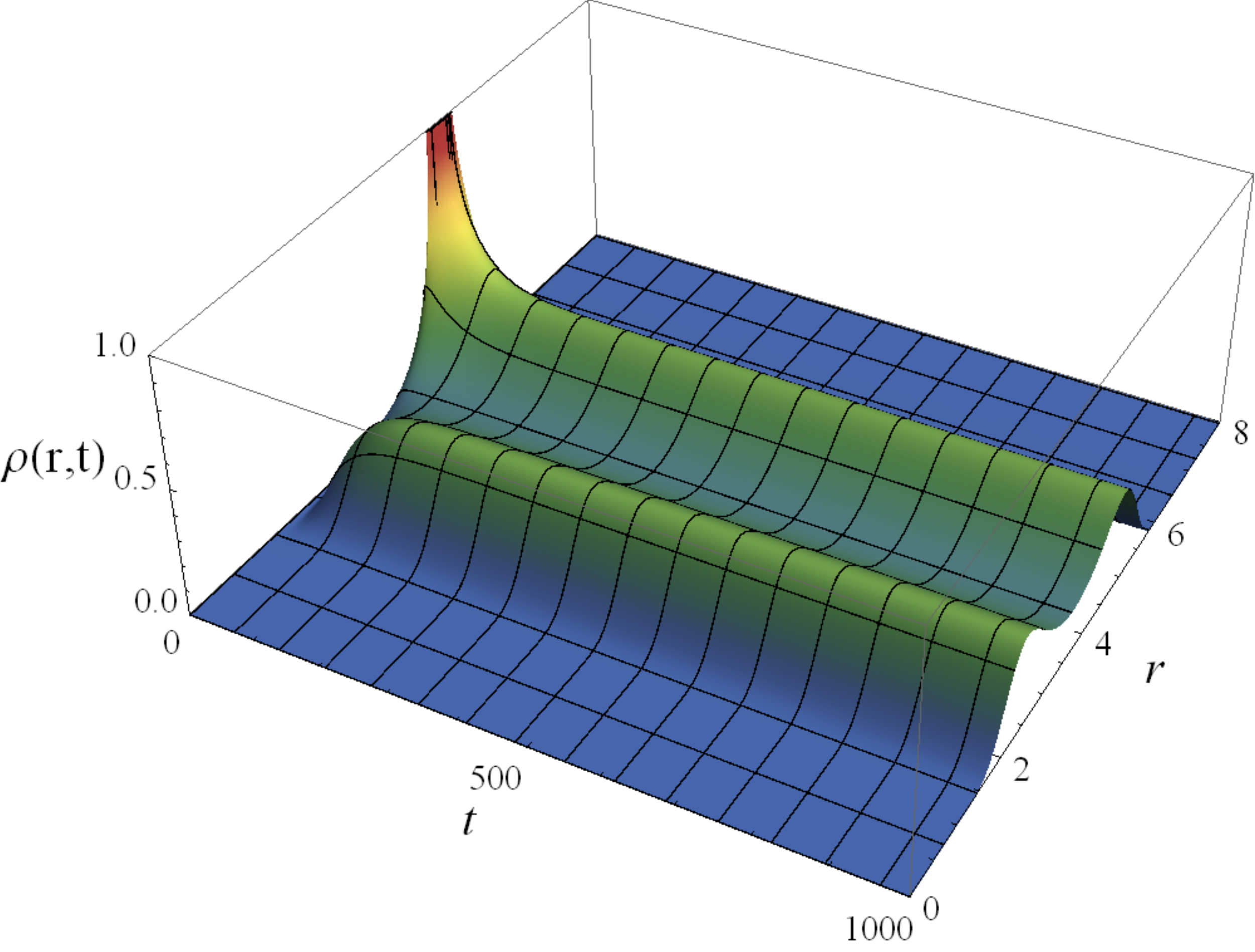}}}
\caption{Time evolution of the probability distribution $\rho(r, t)$ with $\alpha$=2. In top and bottom rows, the Gaussian wave pockets are initially located at the small and large black hole states, respectively. We choose the reflection boundary conditions at $r$=0 and $r$=+$\infty$. (a) $T_{\rm E}$=0.044. (b) $T_{\rm E}$=0.045. (c) $T_{\rm E}$=0.044. (d) $T_{\rm E}$=0.045.}\label{aaThreeDrho45l}
\end{figure}

The time evolution of the probability distribution $\rho(r, t)$ is plotted in Fig. \ref{aaThreeDrho45l}. In Figs. \ref{ThreeDrho44s} and \ref{ThreeDrho45s}, the initial probability distributions, the Gaussian wave packets, locate at the small black hole states while with temperature $T_{\rm E}$=0.044 and 0.045, respectively. At $t$=0, the probability distribution is near the small black hole state. With the evolution of the time, the peak of $\rho(r, t)$ at $r=r_{\rm hs}$ decreases, while another peak locating at the large black hole starts to increase. This implies that the system tends to leak to the large black hole state. Moreover, we observe that the probability distribution reaches a quasi-stationary distribution at a short time, and the peaks are at the small and large black hole states, where the double wells of the Gibbs free energy locate at. At long time limit, $\rho(r, t)$ tends to a final stationary distribution. If we initially place the Gaussian wave packets at the large black hole state, we observe the similar phenomena, see Figs. \ref{ThreeDrho44l} and \ref{ThreeDrho45l}.

\begin{figure}
\center{\subfigure[]{\label{Rho44}
\includegraphics[width=7cm]{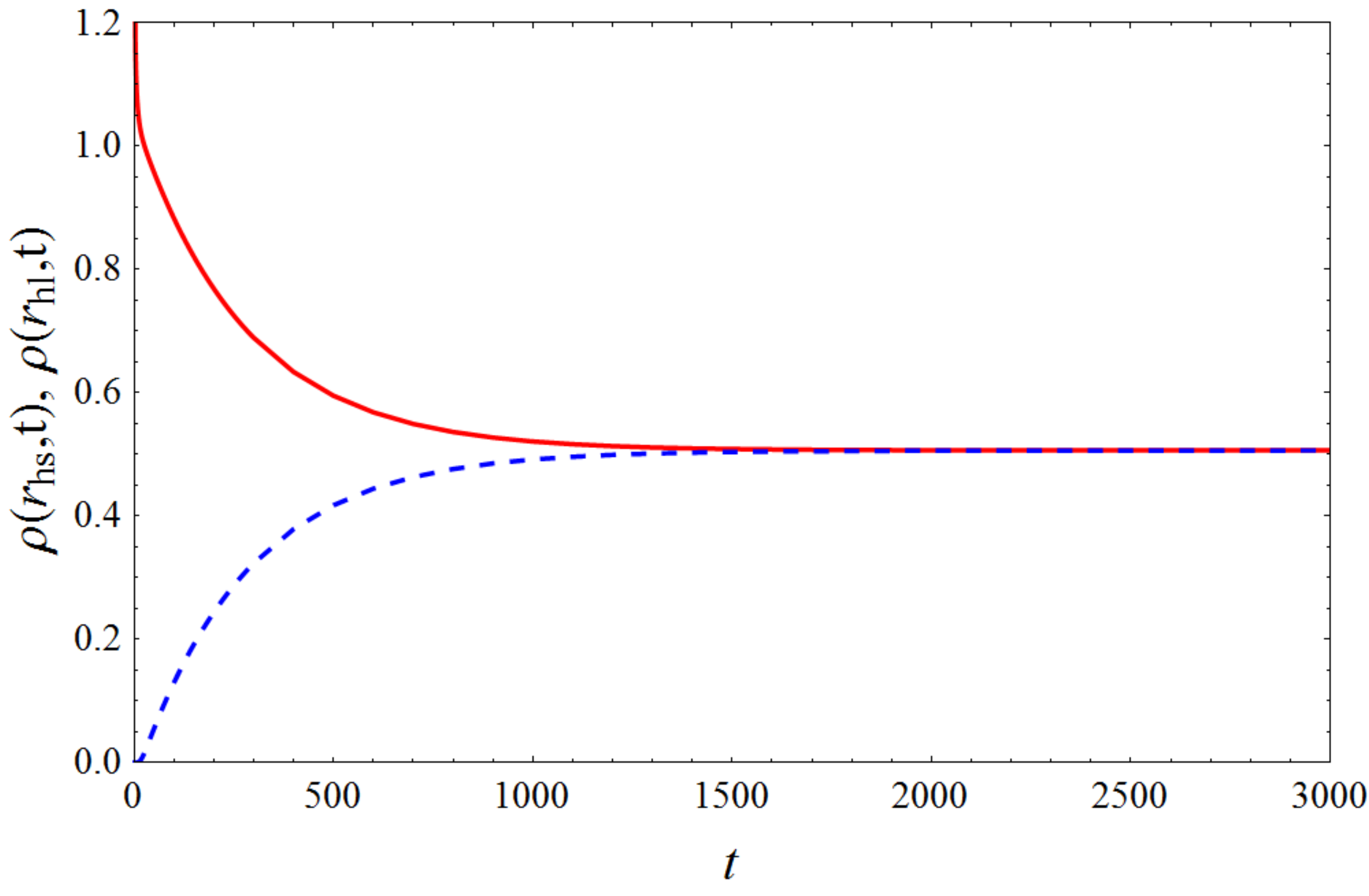}}
\subfigure[]{\label{Rho45}
\includegraphics[width=7cm]{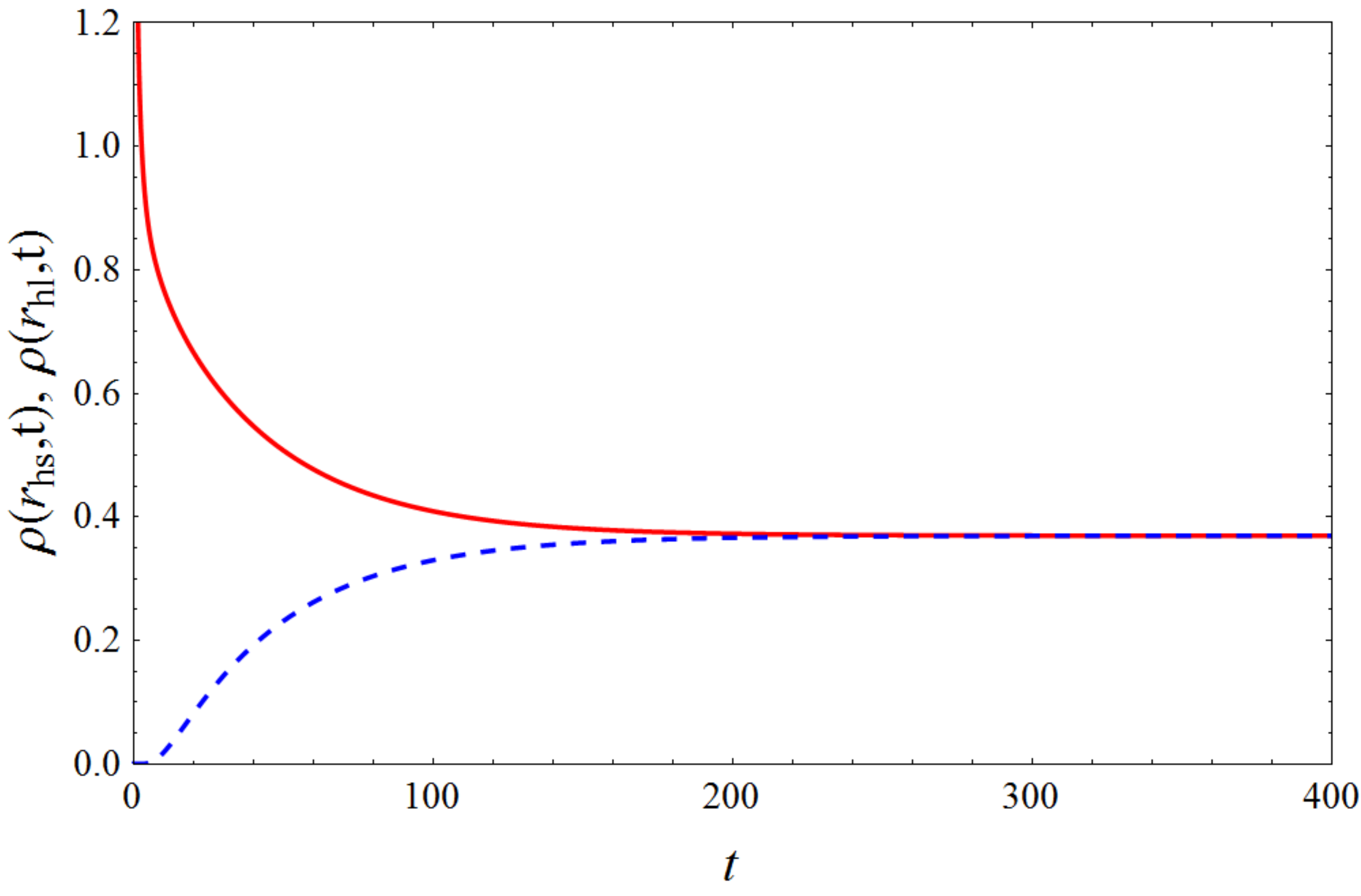}}}
\caption{Time evolution of the probability distribution $\rho(r_{\rm hs}, t)$ (red curves) and  $\rho(r_{\rm hl}, t)$ (blue dashed curves) with $\alpha$=2. The Gaussian wave packets are initially placed at the small black holes. (a) $T_{\rm E}$=0.044. (b) $T_{\rm E}$=0.045.}\label{aRho45}
\end{figure}

In order to make the processes more clear, we plot the probability distributions $\rho(r_{\rm hs}, t)$ and  $\rho(r_{\rm hl}, t)$ in Fig. \ref{aRho45}, for $T_{\rm E}$=0.044 and 0.045, respectively, when the Gaussian wave packets are initially placed at the small black holes. At $t$=0, $\rho(r_{\rm hs}, t)$ takes finite value while $\rho(r_{\rm hl}, t)$ vanishes. This is consistent with that the Gaussian wave packets are at small black hole state at first. Then with the increase of the time, $\rho(r_{\rm hs}, t)$ decreases, while $\rho(r_{\rm hl}, t)$ increases. This indicates the system leak from the small black hole state to the large black hole state. At long time limit, these probability approaches to the same values for each temperature. For examples, $\rho(r_{\rm hs}, t)$ and $\rho(r_{\rm hl}, t)$ tend to 0.5063 and 0.3693 for $T_{\rm E}=0.044$ and 0.045, respectively. The reason for this is that after a long time evolution, the probability distribution approaches a final stationary, and then $\rho(r, t)=\rho(r)$. This leads to the left side of (\ref{FPeq}) vanishes. After the derivation, one can find $\rho(r)\propto e^{-\beta G_{\rm L}(r)}$, which means that the final stationary distribution is only dependent of the Gibbs free energy $G_{\rm L}$, while independent of the initial conditions.

\begin{figure}
\center{\subfigure[]{
\includegraphics[width=7cm]{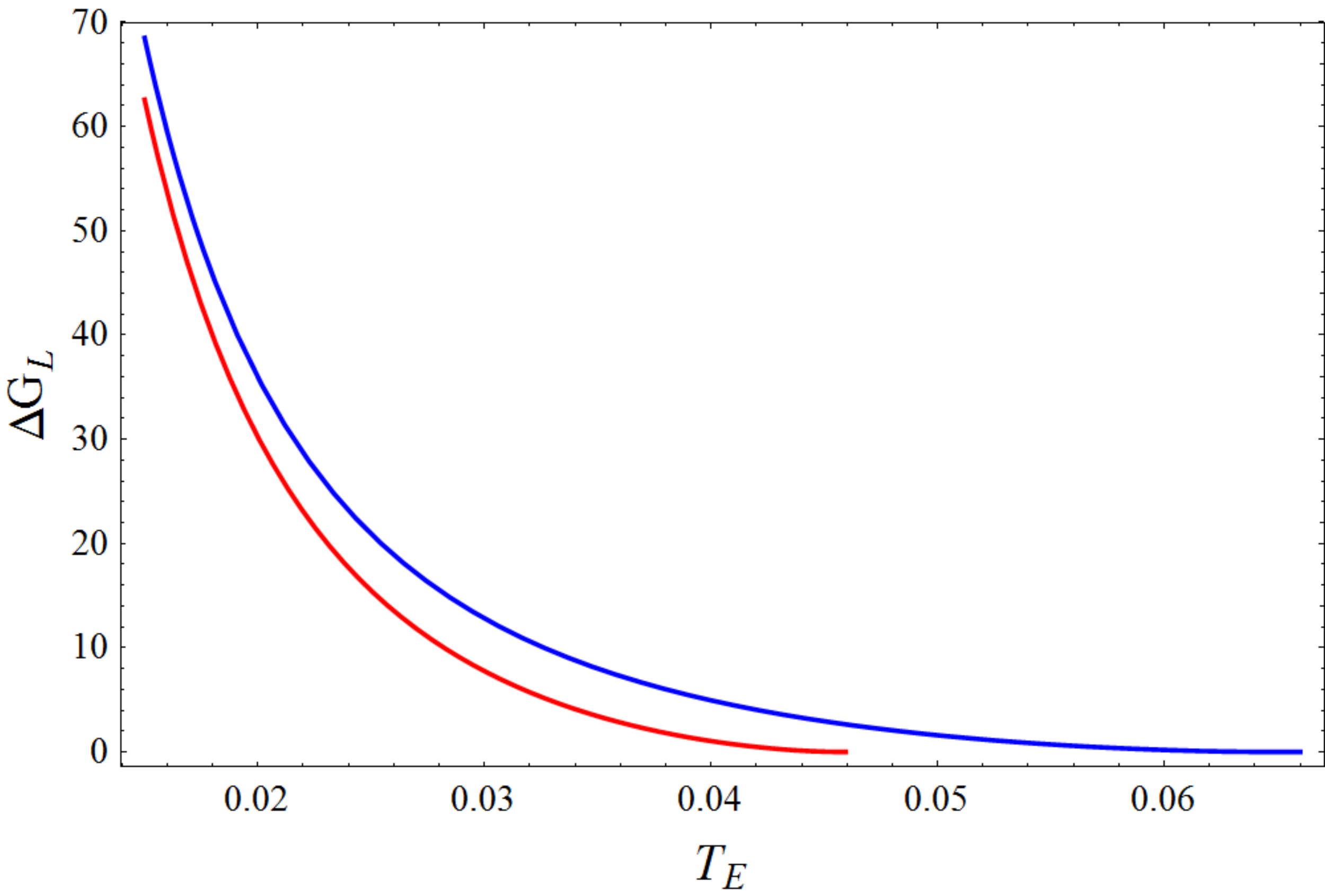}}}
\caption{The height of the barrier $\Delta G_{\rm L}$ as a function of the temperature $T_{\rm E}$ with $\alpha$=1 (top blue curve) and $\alpha$=2 (bottom red curve).}\label{DeltaG}
\end{figure}

Another interesting phenomenon we observe from Fig. \ref{aRho45} is that $\rho(r, t)$ approaches the final stationary at $t$=2000 for $T_{\rm E}=0.044$ and $t$=250 for $T_{\rm E}=0.045$. So this suggests that the system gets a stationary state more rapidly for a higher temperature system. Actually, this originates from the decrease of the barrier of $G_{\rm L}$. After a simple calculation, we obtain the height of the barrier $\Delta G_{\rm L}$
\begin{eqnarray}
 \Delta G_{\rm L}=\frac{27\pi \alpha \left(48\pi^2\alpha T_{\rm E}^2+\sqrt{9-192\pi^2\alpha  T_{\rm E}^2}-3\right)^2}{4\left(3-\sqrt{9-192\pi^2\alpha T_{\rm E}^2}\right)^3}.
\end{eqnarray}
Its behavior is described in Fig. \ref{DeltaG} for fixed $\alpha$=1 and 2. Clearly, $\Delta G_{\rm L}$ decreases with $T_{\rm E}$ and vanishes at the critical temperature. Therefore, for the small-large black hole phase transition, a high temperature system has the fastest relaxation time for the system to approach its final stationary state. Moreover, for a low fixed temperature, $\Delta G_{\rm L}$ has a larger value for small $\alpha$, which indicates that the system of large $\alpha$ has a fast relaxation time. When $T_{\rm E}$ is larger than the critical temperature, there will be no the small-large black hole phase transition, and thus the picture greatly changes.

\subsection{First passage time}

Here we would like to consider the first passage process for the black hole phase transition. The first passage time is defined as the time that the present state of the black hole, for example the small or large black hole state located at the well of $G_{\rm L}$, to reach the intermediate transition state located at the barriers of $G_{\rm L}$ for the first time.

We denote $\Sigma(t)$ as the probability that the present state of black hole has not made a first passage by time $t$, which is given by
\begin{eqnarray}
 \Sigma(t)=\int_0^{r_{\rm m}}\rho(r, t)dr,\label{s1}
\end{eqnarray}
where
\begin{eqnarray}
 r_{\rm m}=\frac{24\pi \alpha T}{3-\sqrt{9-192\pi^2\alpha T^2}}.
\end{eqnarray}
By making use of it, the first passage time $F_{\rm p}(t)$ can be expressed as
\begin{eqnarray}
 F_{\rm p}(t)=-\frac{d\Sigma(t)}{dt}.\label{s2}
\end{eqnarray}
Obviously, $F_{\rm p}(t)dt$ measures the probability that a small black hole state passes through the intermediate black hole state for the first time in the time interval ($t$, $t+dt$). After the first passage, the black hole state will leave the system. Then with the time, the probability that the system stays at the small black hole state decreases and approaches to zero when $t\rightarrow+\infty$.

Substituting Eq. (\ref{s1}) into Eq. (\ref{s2}), and using the Fokker-Planck equation (\ref{FPeq}), one has \cite{LiWang}
\begin{eqnarray}
 F_{\rm p}(t)&=&-\frac{d}{dt}\int_{0}^{r_{\rm m}}\rho(r,t)dr\nonumber\\
       &=&-\int_{0}^{r_{\rm m}}\frac{\partial}{\partial t}\rho(r, t)dr\nonumber\\
       &=&-D\int_0^{r_{\rm m}}\frac{\partial}{\partial r}
       \left(e^{-\beta G_{\rm L}(r)}\frac{\partial}{\partial r}\left(e^{\beta G_{\rm L}(r)\rho(r, t)}\right)\right)dr\nonumber\\
       &=&-De^{-\beta G_{\rm L}(r)}\frac{\partial}{\partial r}\left(e^{\beta G_{\rm L}(r)}\rho(r, t)\right)\bigg|_{0}^{r_{\rm m}}\nonumber\\
       &=&-D\frac{\partial\rho(r, t)}{\partial r}\bigg|_{r_{\rm m}}.
\end{eqnarray}
where at $r$=0 and $r=r_{\rm m}$, we have imposed the reflecting boundary condition and the absorbing boundary condition, respectively. If considering the transition from the large black hole state to small black hole state, we should impose the reflecting boundary condition at $r=\infty$ instead at $r$=0. And then the first passage time is $F_{\rm p}(t)=D\frac{\partial\rho(r, t)}{\partial r}\big|_{r_{\rm m}}$.

\begin{figure}
\center{\subfigure[]{\label{Rhoab44s}
\includegraphics[width=7cm]{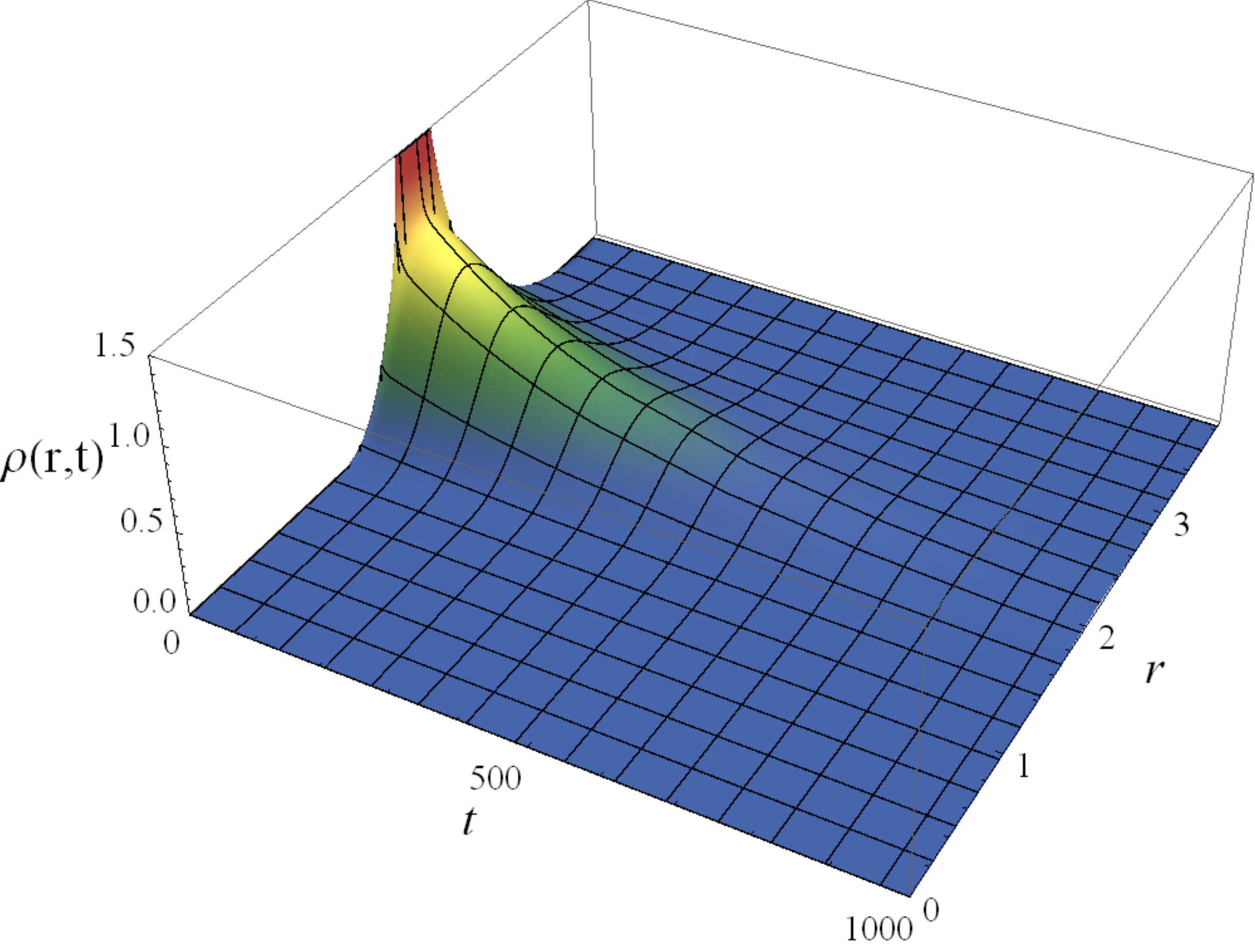}}
\subfigure[]{\label{Rhoab45s}
\includegraphics[width=7cm]{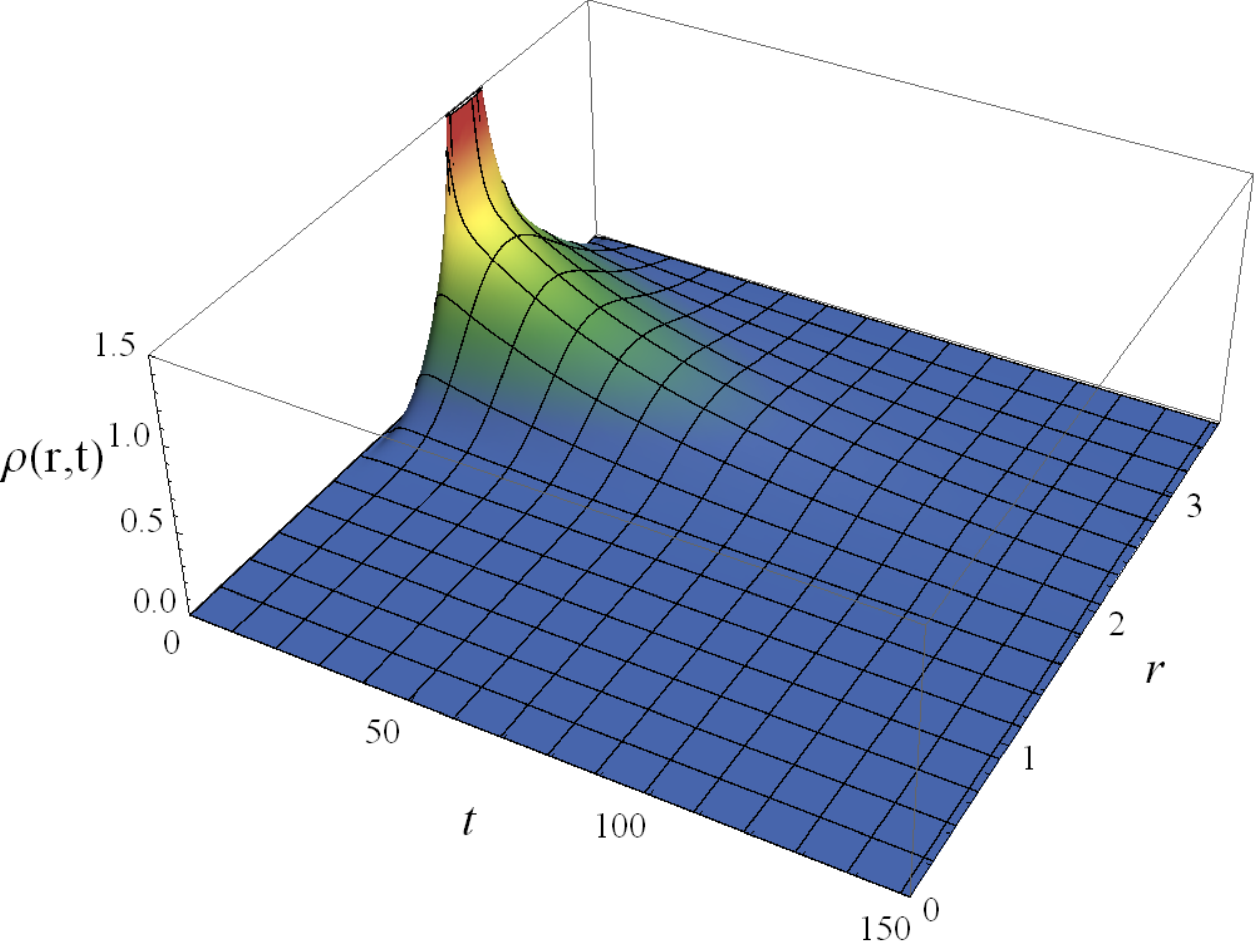}}
\subfigure[]{\label{Rhoab44}
\includegraphics[width=7cm]{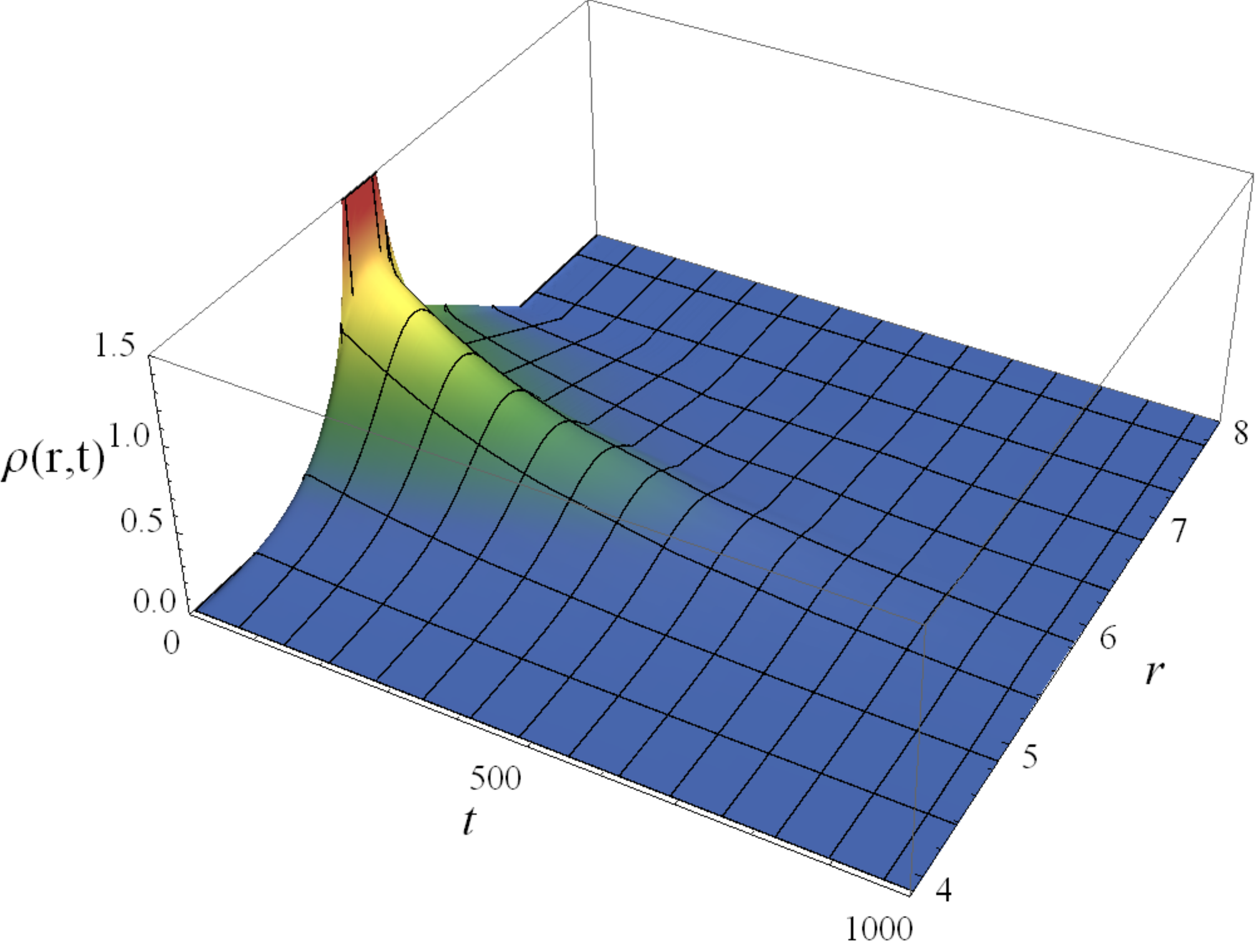}}
\subfigure[]{\label{Rhoab45}
\includegraphics[width=7cm]{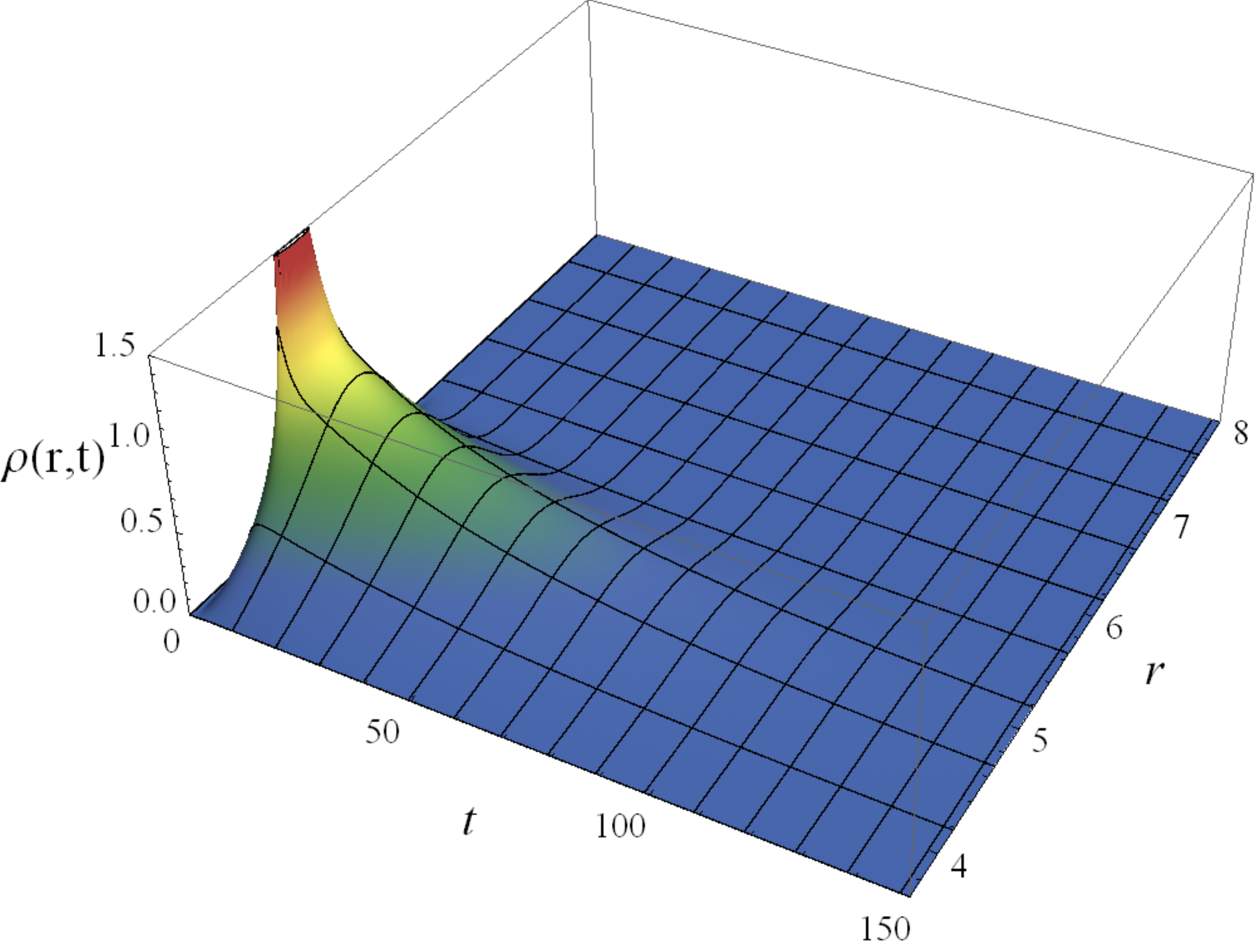}}}
\caption{Time evolution of the probability distribution $\rho(r, t)$ with $\alpha$=2. In top and bottom rows, the Gaussian wave packets are initially located at the small and large black hole states. We choose the reflection boundary conditions at $r$=0 or $r$=+$\infty$ and the absorbing boundary condition at $r$=$r_{\rm m}$. (a) $T_{\rm E}$=0.044. (b) $T_{\rm E}$=0.045. (c) $T_{\rm E}$=0.044. (d) $T_{\rm E}$=0.045.}\label{aaRhoab45}
\end{figure}

\begin{figure}
\center{\subfigure[]{\label{Sigss}
\includegraphics[width=7cm]{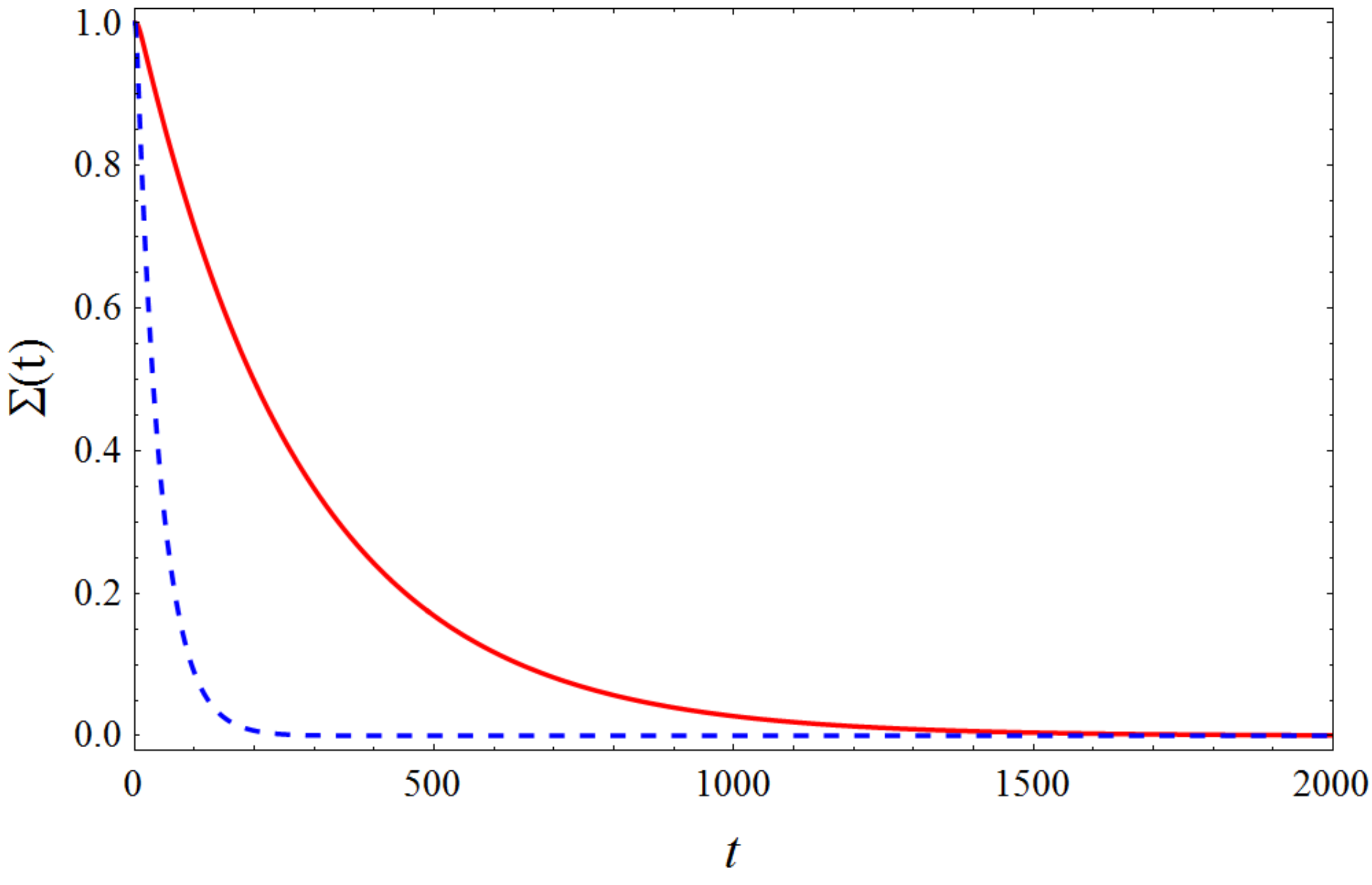}}
\subfigure[]{\label{Sigll}
\includegraphics[width=7cm]{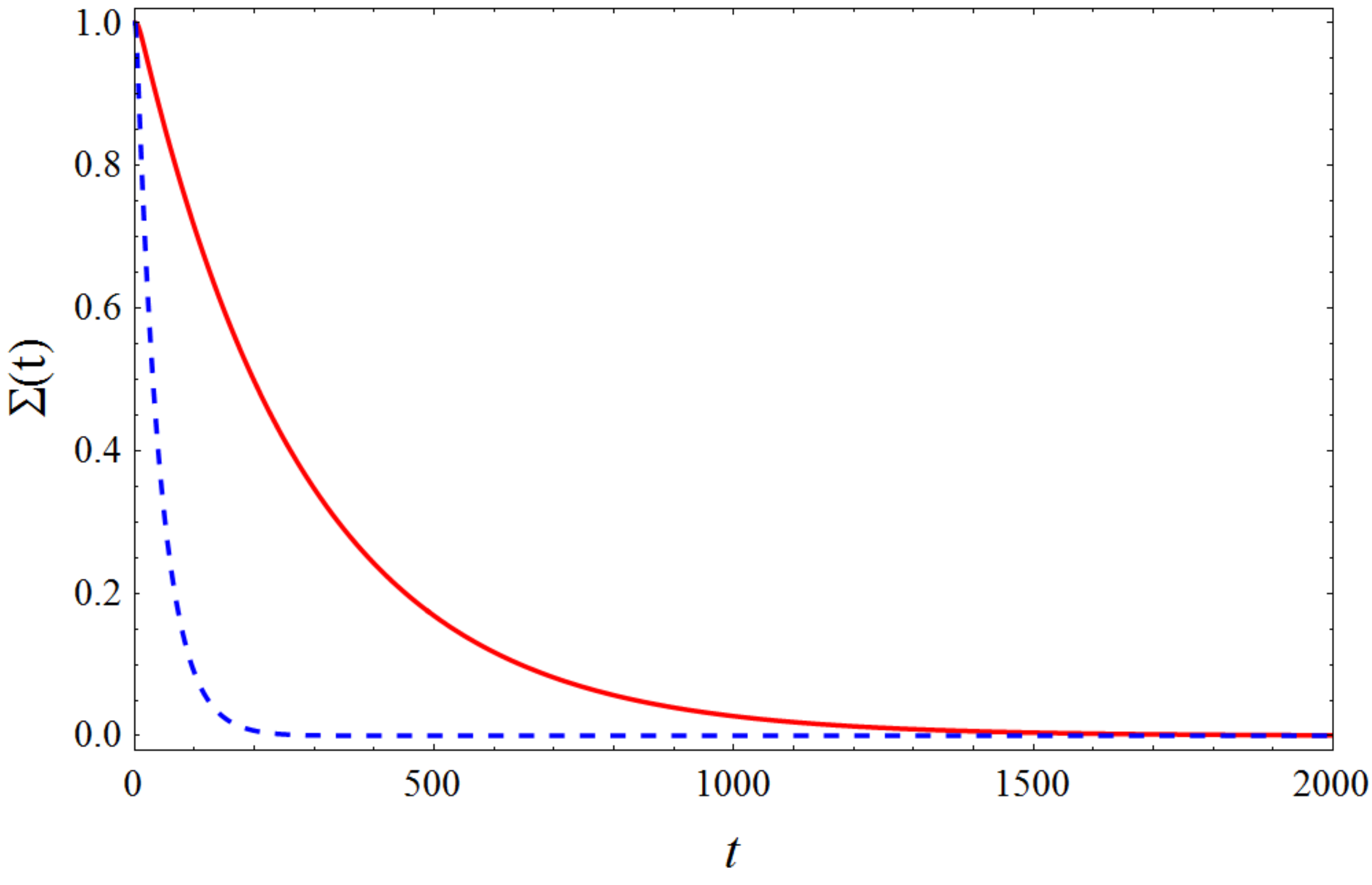}}}
\caption{Time evolution of the probability $\Sigma(t)$ that the system stays at the small black hole state (a) or large black hole state (b). Red solid curves are for $T_{\rm E}$=0.044 and blue dashed curves for $T_{\rm E}$=0.045 with $\alpha$=2.}\label{aSigll}
\end{figure}

First, we numerically solve the Fokker-Planck equation (\ref{FPeq}). Time evolution of the probability distribution $\rho(r, t)$ is plotted in Fig. \ref{aaRhoab45} for the temperature $T_{\rm E}$=0.044 and 0.055, respectively when the Gaussian wave packets are initially located at the small and large black hole states. From these figures, it is clear that the wave packets decay quickly and disappear at short time. In order to show the decreases of the probability $\Sigma(t)$ for the system staying at the small or large black hole state, we display the probability $\Sigma(t)$ in Fig. \ref{aSigll}. Obviously, $\Sigma(t)$ will not be preserved due to that the absorbing boundary condition is imposed. For both the cases that the small black hole state transits to the large black hole state and the reversed case, $\Sigma(t)$ rapidly decreases with the time. Moreover, the higher the temperature is, the faster the probability decreases.

\begin{figure}
\center{\subfigure[]{\label{Fptl44}
\includegraphics[width=7cm]{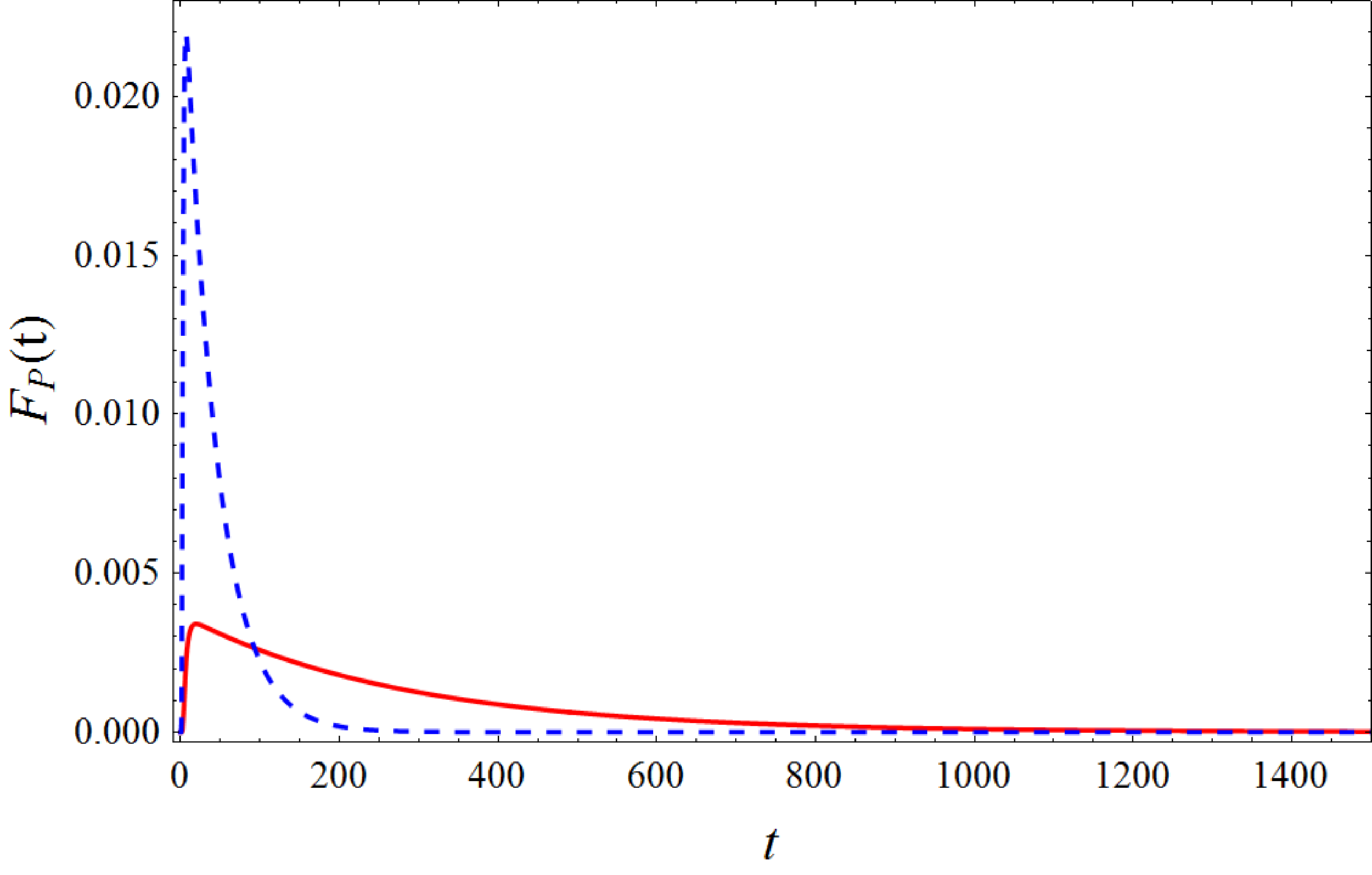}}
\subfigure[]{\label{Fptl45}
\includegraphics[width=7cm]{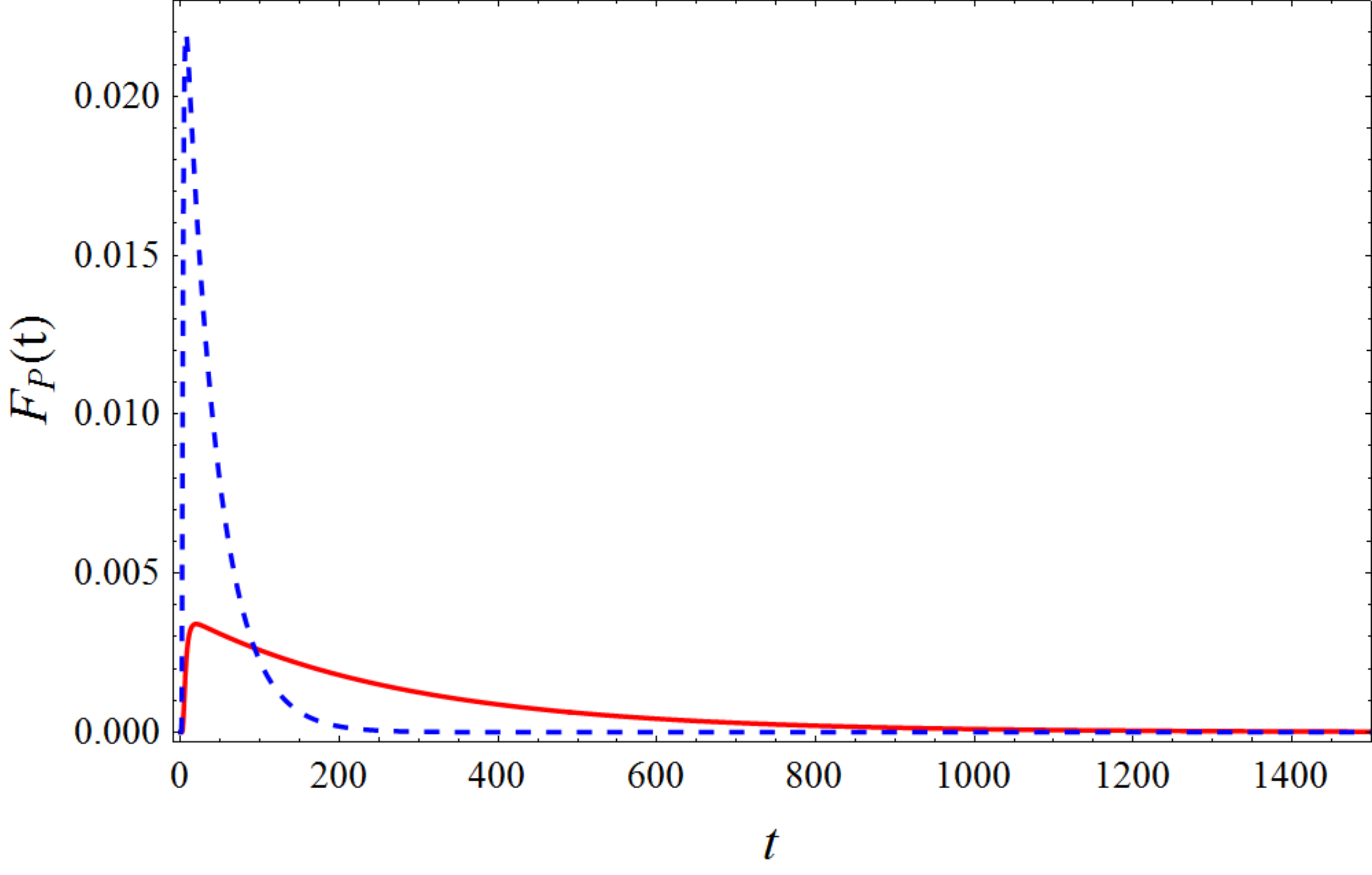}}}
\caption{The distributions of the first passage time $F_{\rm p}(t)$. Red solid curves are for $T_{\rm E}$=0.044 and blue dashed curves for $T_{\rm E}$=0.045 with $\alpha$=2. (a) From small black hole to large black hole. (b) From large black hole to small black hole.}\label{aFptl45}
\end{figure}

The distributions of the first passage time $F_{\rm p}(t)$ are displayed in Fig. \ref{aFptl45}. These two cases that from small to large black holes and from large to small black hole, the behavior of $F_{\rm p}(t)$ is the similar. For each temperature, there exists a single peak at short time. This suggests that a considerable fraction of the first passage events takes place at small $t$ before $F_{\rm p}(t)$ approaches its exponential decay form. Furthermore, with the increase of the temperature, the peak increases and is shifted to the left. The most reason is because that the height of the barrier decreases with the temperature, see Fig. \ref{DeltaG}, which makes the transitions more easier to occur.

\section{Conclusions and discussions}
\label{Conclusion}

In this paper, we studied the dynamic process of the stable small-large black hole thermodynamic phase transition for the five-dimensional neutral GB AdS black holes on the Gibbs free energy landscape.

At first, we briefly reviewed the thermodynamic properties for the black hole. Making use of the equation of state, this black hole admits a small-large black hole phase transition. Especially, there exists an analytical formula for the coexistence small and large black holes, which provides us a good chance to exactly study the black hole phase transition. Employing the analytical coexistence curve, we showed that the horizon radius has a sudden change among the phase transition. Moreover, this change at the critical point has a critical exponent $\frac{1}{2}$. Thus the change of the horizon radius acts as an order parameter, which can be used to characterize the small-large black hole phase  transition for the GB AdS black holes. This also inspires us to consider the dynamic process of the phase transition, for example, the Fokker-Planck equation, by coordinating with the horizon radius.

Since the free energy plays an important role on determining the black hole phase transition point, we examined the classical swallow tail behavior of the Gibbs free energy $G$. On the other hand, on the Gibbs free energy landscape, there is a new one free energy $G_{\rm L}$. Generally, these black holes in this landscape are not the real solution of the field equations unless the temperature of the ensemble equals the Hawking temperature. By using the first law of black holes, we proved that on this Gibbs free energy landscape, the real black holes always locate at the extremal point of $G_{\rm L}$. More importantly, the local maximal points correspond to thermodynamic unstable black holes, which have negative heat capacity. Meanwhile, the local minimal points denote the stable black holes. If there are two wells of $G_{\rm L}$. The system prefers the one with the lower Gibbs free energy, while the other one is a global unstable state, i.e., the metastable state. It is worth to point out that in Ref. \cite{LiWang}, the authors considered the dynamic process of a phase transition between metastable and stable black holes. For the case of a stable small-large black hole phase transition, these two wells of $G_{\rm L}$ must have the same depth. And in this paper, we focused on this specific case.

After clarifying the certain properties of the Gibbs free energy on the landscape, we turned to consider the dynamic process governed by the Fokker-Planck equation of the small-large black hole phase transition. After imposing the reflecting boundary conditions at $r$=0 and $+\infty$, we numerically solved the Fokker-Planck equation when place the Gaussian wave packets at small and large black hole states, respectively. The results indicate that the probability will leak to another state from the initial state with the time. After a long time evolution, the probability gets a stationary distribution. And the probability at the small and large black hole equals, which is because that these two black holes have the same Gibbs free energy. Significantly, higher temperature system has a shorter time to approach its stationary distribution.

The first passage process was also considered. Imposing the intermediate transition state with the absorbing boundary condition, we resolved the Fokker-Planck equation. Since the initial state will leave after the passage events, so the probability decreases with the time. This was confirmed in Fig. \ref{aSigll}. After a long time evolution, the probability that stays at the initial state vanishes. The higher the temperature is, the faster the probability decreases. The first passage time was also calculated. For each case, there presents a single peak indicating a considerable fraction of the first passage events take place at short time.

In summary, we firstly examined the dynamic process of the stable small-large black hole phase transitions in a modified gravity. The results reveal some specific properties of the process. As we know, there exist more interesting phase transitions, such as the reentrant phase transition and triple point, in GB gravity and other gravities, so it is worth to generalize the study to other novel phase transitions. These we expect will uncover the intriguing dynamic properties of thermodynamic phase transition in modified gravities.

As an example, here we would like to address how to generalize the study for a neutral GB-AdS black hole to a charged one. As have shown previous studies, the small-large black hole phase transition is universal for the charged GB-AdS black hole in four or high dimensions. So in addition to GB coupling $\alpha$, the phase transition and critical point will also rely on the black hole charge. Although an extra charge parameter is included in, the pattern of $G_{\text{L}}$ will not change. So the treatment of this paper can be directly extended to the charged black hole case. The result will uncover the particular property of the charge on the dynamics of the phase transition for different GB coupling parameter $\alpha$.

\section*{Acknowledgements}
This work was supported by the National Natural Science Foundation of China (Grants No. 12075103, No. 11675064, No. 11875151, and No. 12047501), the 111 Project (Grant No. B20063), and the Fundamental Research Funds for the Central Universities (No. Lzujbky-2019-ct06).

\end{document}